\documentclass[a4paper]{panl}
\usepackage{cite}
\usepackage{wrapfig}
\usepackage{graphicx}
\usepackage{amssymb}
\usepackage{amsfonts}
\usepackage{amsmath}
\usepackage{longtable}
\usepackage{rotating}
\usepackage{lscape}
\usepackage{epsfig}
\usepackage{multirow}
\usepackage{makecell,tabularx}
\originalTeX

\begin{document}

\title{Monte-Carlo Generator of Heavy Ion Collisions DCM-SMM \\ }
\maketitle
\authors{M.\,Baznat$^{a,}$,
A.\,Botvina$^{b,d,}$,
G.\,Musulmanbekov$^{c,}$\footnote{Corresponding author E-mail: genis@jinr.ru},
V.\,Toneev$^{c,}$,
V.\,Zhezher$^{c,}$\\}

\from{$^{a}$\,IAP, Kishineu, Moldova}
\vspace{-3mm}
\from{$^{b}$\,INR, RAS, Moscow, Russia}
\vspace{-3mm}
\from{$^{c}$\,JINR, Dubna, Russia}
\vspace{-3mm}
\from{$^{d}$\,ITP and FIAS, University of Frankfurt am Main, Germany}

\begin{abstract}
The new monte-carlo generator of heavy ion collisions, DCM-SMM, based on Dubna Cascade Model (DCM-QGSM) and Statistical Multifragmentation Model (SMM) is described. The model aimed to generate particle--nucleus and nucleus--nucleus collisions at a wide range of energy was created to provide the computer simulation support to new experimental facilities BMN and MPD at the accelerator complex NICA. It can simulate the production of both light particles and nuclear fragments and hyperfragments on the event by event basis. \\
\vspace{0.2cm}

\end{abstract}
\vspace*{6pt}


\label{sec:intro}
\section*{\bf Introduction}
Modern experiments at heavy ion facilities require simulation at all stages of their planning, construction and functioning. The important role in this process belongs Monte-Carlo models and their computer codes of nuclear collisions. Monte-Carlo simulation is very effective tool for optimizing the detector elements, debugging the event reconstruction algorithms, predicting the efficiency, calculating the signal-to-background ratio, determining the best criteria for selecting events.
In data analysis, on the other hand, the model must provide, first of all, with a good background for (un)expected effects including (as much as possible) all mechanisms describing the properties of products of reactions and the various effects understandable in the framework of modern theories.
Study of the properties of strongly interacting matter in heavy ion collisions is the main task of the current and future experiments over the world. The theory of strong interactions, QCD, predicts that the nuclear matter may convert in such collisions into a new, QGP state. It is necessary to have reliable models and codes including a wide variety of heavy ion related effects ranging from particle production, hypernuclei formation and multifragmentation to correlations and collective flow. To study the possibilities of forming new states in a hot and dense nuclear matter the new experimental facilities BMN and MPD at the new heavy ion collider NICA are being created. Needless to say that these experiments require reliable transport generators. To meet these requirements the new transport model DCM-SMM, Dubna Cascade - Statistical Multifragmentation Model, for simulation of products of reactions in heavy ion collisions in the energy range from hundred MeV to hundred GeV is created. The basic components of the DCM-SMM are the Dubna Cascade Model (DCM) \cite{barash,toneev83}, the Quark-Gluon String Model (QGSM) \cite{toneev90,amelin51a,amelin51b,amelin52} and the Statistical Multifragmentation Model (SMM) \cite{SMM-3}. New physics phenomena are implemented in the model: extended coalescence, multifragmentation, hyperfragments production, vorticity of nuclear matter and Lambda polarization. Currently the model  Accordingly, the paper is organized as follows: Section 1 starts with a brief description of components of the model DCM and QGSM. New coalescence model of formation of light and medium clusters and hyperon clusters from secondary particles in a wide rapidity interval is described in Section 2. Section 3 is devoted to light and intermediate mass fragment  production by excited residual nuclei. Analysis of possibility of forming a nuclear vorticity field in non-central heavy ion collisions resulting in a global polarization of lambda-hyperon is given in Section 4. Section 5 demonstrates comparison of the model with experimental data. In Section 6 we discuss the results and further improvements of the model. The brief guide for running the program code with an example is given in Appendix.

\label{sec:DCM-QGSM}
\section{\bf Dubna Cascade Model and Quark Gluon String Model , DCM-QGSM}
One of the first models designed to describe the dynamics of energetic heavy-ion collisions was the intra-nuclear cascade model developed in Dubna
\cite{toneev83}. The Dubna Cascade Model, DCM, is based on the Monte-Carlo solution of a set of the Boltzmann-Uehling-Uhlenbeck relativistic kinetic equations with the collision terms, including cascade-cascade interactions.  The modified non-Markovian relativistic kinetic equation, having a structure close to the Boltzmann-Uehling-Uhlenbeck kinetic equation, but accounting for the finite formation time of newly created hadrons, is used for simulations of relativistic nuclear collisions. Particle-nucleus and nucleus-nucleus collisions are treated as noncoherent superposition of binary interactions. For particle energies below 1~GeV it is sufficient to consider only nucleons, pions and deltas. The model includes a proper description of meson and baryon dynamics for particle production and absorption processes.
The black disk approximation is adopted as criterion of interaction. It means that two hadrons can interact both elastically and inelastically if the
distance $d$ between them is smaller than $\sqrt{\sigma/\pi}$, where
$\sigma$ is the total cross section. Tables of the experimentally
available information, such as hadron cross sections, resonance
widths and decay modes, are implemented in the model.
The model includes the concept of formation time which is defined by uncertainty principle $\tau_{f}\sim \hbar /m_{\bot}$. Formation time, $\tau_{f}$, in turn, defines the length of hadron formation, $l_{f} = \tau_{f} (p/m_{\bot})$.
 For all produced particles the appropriate formation time taken by a reasonable agreement with experimental data.
Nuclei are generated as Fermi gas of nucleons with Wood-Saxon density distribution
\begin{equation}
 \rho(r)=\rho_{0}/{1+exp[(r-r_{0}/a)]},
\end{equation}
with
\begin{equation}
r_{0}=1.19A^{1/3}+1.61A^{-1/3} fm, a=0.54 fm
\end{equation}
To take the Fermi motion of nucleons into account a Fermi momentum $p$ is generated for each nucleon uniformly distributed in the range $0 < p < p_{F}$, where $p_{F}$ is the maximum Fermi nucleon momentum. Fermi distribution of nucleon momenta provides Pauli blocking factors for scattered nucleons. The nuclear potential is treated dynamically, i.e., for the initial state it is determined using the Thomas-Fermi approximation, but later on its depth is changed according to the number of knocked-out nucleons. This allows one to account for nuclear binding.

DCM is a universal intranuclear cascade model to describe lepton, hadron and nucleus-nucleus interactions.  Cascade particles produced in primary binary
 interactions then passage through both the target and projectile nuclei producing in turn new secondary particles.
 The model includes interactions of cascade particles with each other, as well. It uses experimental cross sections  for these elementary
interactions to simulate angular and energy distributions of cascade particles, also considering the Pauli exclusion principle. Cascade particles are traced until their energy decreases due to elementary collisions to a value equal or below the cutoff energy of 1 MeV (plus the Coulomb barrier, for protons) above the Fermi level, when they are considered to be absorbed by the target/bombarding nucleus, increasing its excitation energy. When all of the cascade particles escape from or are absorbed by the target and bombarding nuclei, the fast stage of the reaction is ceased.
Usually the residual nucleus (RN) produced after the completion of the intranuclear cascade is considered as thermalized many-body system. However, the system of Fermi particles formed just after the cascade may be out of equilibrium. In the course of the expansion and equilibration such a system may emit preequilibrium particles \cite{blann}. As a result, the excitation energy and the nucleon content of the primary RN may differ
considerably from the corresponding values for the thermalized system at freeze-out.
 The subsequent relaxation of a residual nucleus at the equilibrium evaporation/fission stage is described by using the generalized (sequential) evaporation model \cite{GEM2}.
Development of DCM is described in papers \cite{toneev90,amelin51a,amelin51b,amelin52}.

To make the DCM code applicable at higher energies (up to hundreds GeV/nucleon), it was merged with the Quark-Gluon String Model (QGSM).
QGSM simulating elementary hadron collisions at the energies higher than about 5~GeV describes binary collisions in the framework
of independent quark-gluon strings quasiclassical approximation \cite{toneev90,amelin51a,amelin51b,qgsm}.
In this treatment, collisions of hadrons lead to the formation of the strings between quark and antiquark and quark and diquark. The
production of new particles occurs via fragmentation
of specific colored objects, strings. Strings are uniformly
stretched, with constant string tension $\kappa\approx 1GeV/fm$, between
the quarks, diquarks, and their antistates. The excited strings then
fragment into pieces via the Schwinger-like mechanism, and the produced hadrons are uniformly distributed in the rapidity space. Hadron production is treated in the framework of Dual Parton Model \cite{capella,kaidalov} which assumes that the main contribution to particle production is due to soft processes composed of elastic, diffractive and non-diffractive interactions. In the QGSM, the leading edges of stretched strings are replaced by energetic hadrons; this corresponds to minimal inclusion of quark dynamics. This means that the basic kinetic equations will be written in terms of hadronic states. However, the quark properties are used for specifying the initial hadron-hadron states and for describing the passage of strings through nuclear matter with subsequent hadronization by introducing the concept of hadron formation time. Due to the uncertainty principle newly produced particles can interact further only after a certain formation time. However, hadrons containing the valence quarks can interact immediately with the reduced cross
section $\sigma=\sigma_{qN}$.

The model is based on the 1/N$_c$ expansion of the amplitude for binary processes where N$_c$ is the
number of quark colours. Different terms of the 1/N$_c$ expansion correspond to different diagrams which are classified according to their topological properties. Every diagram defines how many strings are created  in a hadronic collision and which quark-antiquark or quark-diquark pairs form
these strings. The relative contributions of different diagrams can be estimated within Regge theory, and all QGSM parameters for hadron-hadron
collisions were fixed from the analysis of experimental data. The
break-up of strings via creation of quark-antiquark and diquark-antidiquark
pairs is described by the Field-Feynman method \cite{field78},
using phenomenological functions for the fragmentation of quarks, antiquarks and diquarks into hadrons.
The QGSM takes into account the lowest SU(3) multiplets in mesonic, baryonic and antibaryonic sectors, so interactions between almost 70 hadron species are treated on the same footing. Particles produced by the model are given by Table in Appendix.

\section{\bf Coalescence: Light and medium cluster production}
\label{sec:coales}
According to early version of DCM, after completion of the cascade stage of a reaction, the coalescence model is applied to "create" high-energy d, t, $^3$He, and $^4$He by final state interactions among emitted cascade nucleons \cite{toneev83, Gudima:83a}.
 Energetic light fragments (LF) heavier than $^4$He may be emitted through three mechanisms: Fermi breakup, coalescence and multifragmentation.
In the initial formulation \cite{toneev83} the coalescence model
forms a deuteron from a proton and a neutron produced after the cascade
stage of reaction if their relative momenta are within a sphere of
radius $p_c$ , comparable to the deuteron momentum. The same momentum
criterion have been used to describe formation of tritons, $^3$He, and
$\alpha$-particles. In particular, the parameters $p_c$(d) = 90,
$p_c$(t) = 108, $p_c$($^3$He) = 108, and $p_c$($\alpha$) = 115 (MeV/c)
were adopted to reproduce the experimental data \cite{toneev83,Gudima2012}.
We believe that the spacial coordinates of nucleons should be taken into
account too after all cascade interactions have stopped.
Here we assume that the coalescence criterion used to form the composite
particles includes the proximity of nucleons both in the momentum and
coordinate space in the system of a cluster. The coordinate coalescence
parameters are determined by the relation $r_c$ = $\hbar / p_c$ ,
with the same values of $p_c$. This coalescence procedure was extended
to consider the formations of known light hypernuclei \cite{Gudima2012},
for example, $^3$H$_\Lambda$, $^4$H$_\Lambda$, $^4$He$_\Lambda$. As an
approximation we use the same coalescence parameters for both
conventional fragments and hyperfragments.
Such a mechanism of light
fragments and hyperfragment production will be dominating in the midrapidity
zone of relativistic ion collisions and can be measured with modern
detector facilities at NICA and Nuclotron. Yield of light (hyper)fragments in AuAu/PbPb collisions calculated according to coalescence mechanism is shown in Fig.~\ref{coales}.
\begin{figure}[ht]
\begin{center}
\begin{tabular}{cc}
\includegraphics[width=63mm]{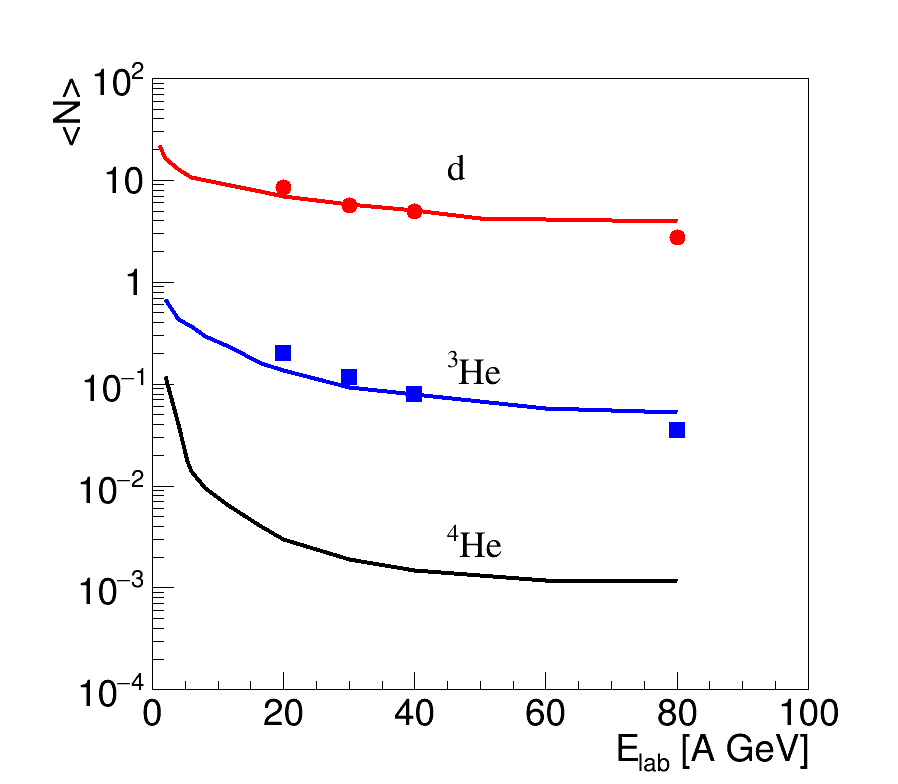}
&
\includegraphics[width=63mm]{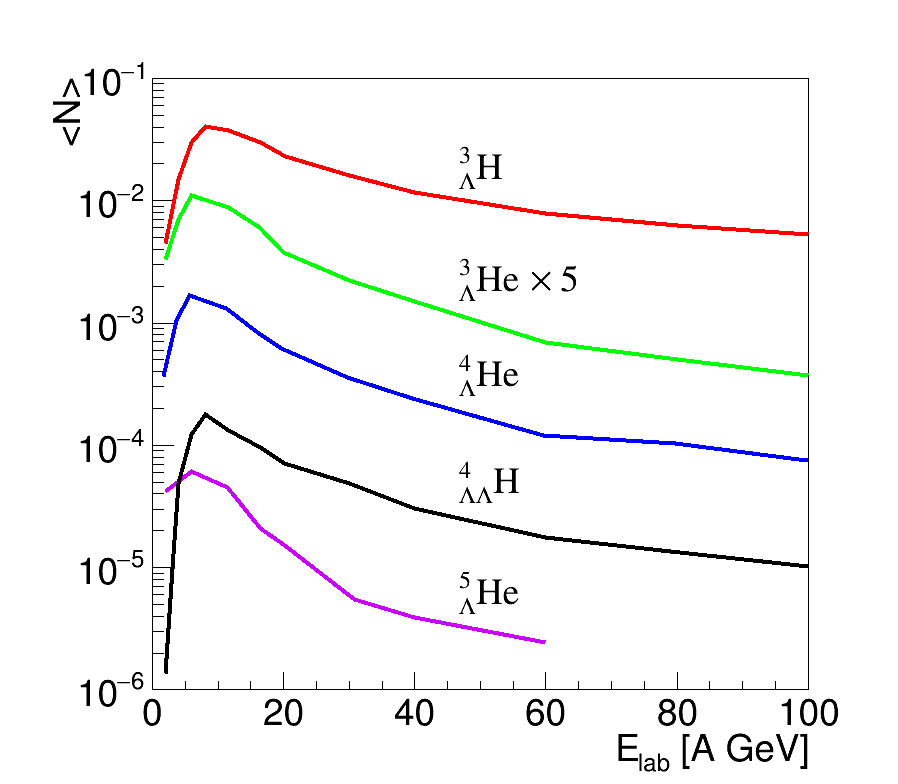}
\end{tabular}
\end{center}
\vspace{-5mm}
\caption{Mean multiplicities of light fragments and hyperfragments formed due to the coalescence mechanism at the AGS and NA49 energy range compared with NA49 data \cite{coal-NA49} on deuteron and $^3$He.}
\labelf{coales}
\end{figure}
Although mean multiplicities of light fragments in the model more or less agree with data, the shapes of their rapidity distributions essentially  differ. As can be seen on Fig.~\ref{d-rap} calculated rapidity distributions of deuterons are concentrated near the center of mass of colliding nuclei. The same is for other light fragments ( Fig.~\ref{He3-rap}). Some of  these deviations may come from nucleon spectra which are concentrated at mid-rapidity at AGS and lower NA49 energies (Figs.~\ref{AGS_prot_y-dist}, \ref{NA49_prot_y-dist}). Another reason of this deviation could be collective effects in
motion of nucleons which are not included presently in DCM. For example,
it could be a hydrodynamical-like expansion of hot matter
produced in the midrapidity region. In this case we expect to get a more
broad nucleon distribution in the rapidity. As well as the transverse
momenta of baryons in this region will increase, and this would lead to
a local decrease of the cluster formation within the coalescence picture
at the midrapidity.
\begin{figure}[ht]
\begin{center}
\includegraphics[width=130mm]{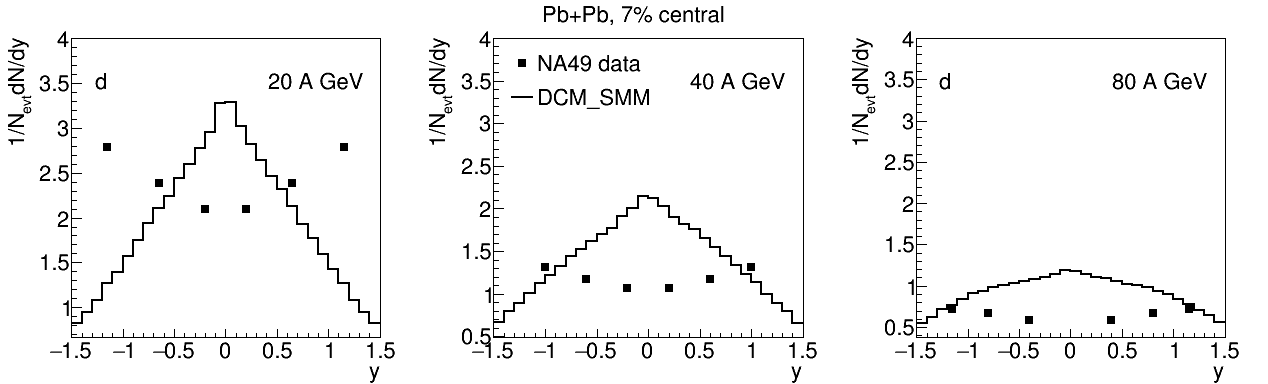}
\end{center}
\vspace{-5mm}
\caption{Rapidity distributions of colalesced deuterons compared with NA49 data \cite{coal-NA49}.}
\label{d-rap}
\end{figure}
\begin{figure}[ht]
\begin{center}
\includegraphics[width=130mm]{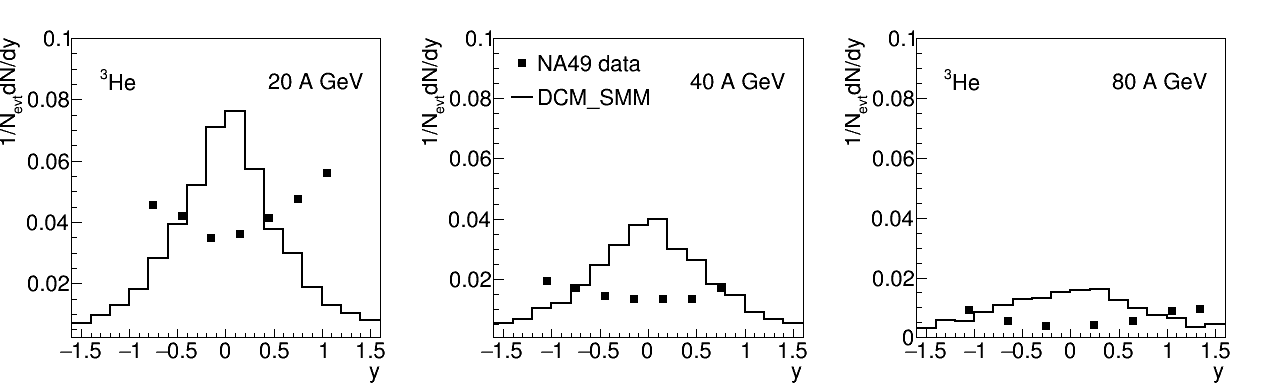}
\end{center}
\vspace{-5mm}
\caption{Rapidity distributions of coalesced $^3$He compared with NA49 data \cite{coal-NA49}.}
\label{He3-rap}
\end{figure}

The important advantage of our coalescence procedure is that it gives
a possibility to analyze the fragment formation on event by event basis,
in particular, by taking into account the correlation with other
particles produced during the cascade stage. This is impossible to
perform in the case of the analytical formulation of the coalescence
model which use the particle spectra integrated over many events.
In future, we plan to adopt a new coalescence approach by considering the
production of coalescent clusters of all sizes \cite{Botvina2015}, We
believe it should also include the formation of exotic and
nuclear/hypernuclear clusters in excited states with their following de-excitation.


\section{\bf Nuclear fragments production}
\label{sec:fragments}
It is well known that highly-excited ($\approx 5-10$ MeV/nucleon) residual nuclei (RN) are produced in inelastic nuclear reactions induced by intermediate- and high-energy particles and nuclei. In these reactions one deals with several stages which differ by characteristic time scales and realized physical conditions. One can distinguish at least three stages: (1) the initial non-equilibrium stage leading to the production of an excited nuclear system; (2) the formation of fragments and break-up of the system into separate fragments; (3) farther  de-excitation of hot fragments via evaporation/fission. The first stage is simulated by intranuclear cascade models. Disintegration of excited residues at the second stage can be described by a wide variety of models that have been proposed for nuclear multifragmentation.
The existing models can be grouped into several categories: probabilistic, macroscopic, statistical models of different kinds, sequential evaporation, and many other models. The previous version of the model, DCM-QGSM includes a preequilibrium stage on which RN with the large excitation energy emits light fragments before transition to the thermalization stage. The excited thermalized RN decays then according to fission and/or sequential evaporation model.

Statistical approaches have proved to be very successful for description of fragment production in nuclear reactions.
According to the statistical hypothesis, initial dynamical
interactions between nucleons lead to re-distribution of the available
energy among many degrees of freedom, and the nuclear system
evolves towards equilibrium. The most famous example of such an
equilibrated nuclear source is the 'compound nucleus' introduced by Niels
Bohr in 1936 \cite{Bohr}.
It was clearly seen in low-energy nuclear reactions leading to excitation
energies of a few tens of MeV. It is remarkable that the statistical
concept works also for nuclear reactions induced by particles
and ions of intermediate and high energies, when nuclei break-up into many
fragments (multifragmentation) \cite{SMM-3}.
In the framework of our combined code, DCM-SMM,  fragment production is subdivided into three stages: (1) a dynamical stage leading to
formation of equilibrated nuclear system, which is described by DCM, (2) disassembly of the system into individual primary fragments described by SMM,
(3) de-excitation of hot primary fragments according to evaporation/fission models. If on the stage 2 we obtain the compound nucleus, then its disintegration takes place at the stage 3 as in the case of other hot fragments.

\subsection{\bf Formation and break-up of thermalized nuclear residues~~}

The DCM was the first model used for realistic calculations of ensembles of
highly excited residual nuclei which undergo multifragmentation, see e.g.
\cite{Botvina90,Botvina94}.
Many dynamical models have also been used for dynamical simulations of ion
reactions, and all models confirm that the character of the dynamical
evolution changes after a few rescatterings of incident nucleons, when high
energy particles ('participants') leave the system.
The time needed for equilibration and transition to the statistical
description is estimated around or less than 100 fm/c for nuclear spectator
matter. Parameters of the predicted equilibrated sources, i.e. their
excitation energies, mass numbers and charges vary significantly depending
on the impact parameter. However, the theoretical calculations and the
analyses of experimental data gives evidences for the saturation of the
spectator residues excitation energy and for an universal connection
between sizes of the residues and their excitation energies
\cite{ALADIN,Xi97,Ogul11,Botvina17}.

\subsection{\bf Evolution from sequential decay to simultaneous break-up~}

After dynamical formation of a thermalized source, its further evolution
depends crucially on the excitation energy and mass number.
The standard compound nucleus picture is valid only at low excitation
energies when sequential evaporation of light particles and fission are
the dominant decay channels \cite{SMM-3,Eren2013}.
However, the concept of the compound nucleus cannot be applied at high
excitation energies, $E^* \geq$ 3 MeV/nucleon. In this case there will
be not enough time for the residual nucleus to reach equilibrium between
subsequent emissions, since the time intervals between subsequent fragment
emissions become very short, of order of a few tens of fm/c \cite{jandel}.
Moreover, the produced fragments will be in the vicinity of each other and,
therefore, should interact strongly. Many theoretical calculations
predict that the compound nucleus will be unstable at high temperatures,
and a simultaneous break-up into many fragments is the only possible way for
the evolution of highly-excited systems \cite{XXXHFTF}.
The rates of the particle emission calculated as for an isolated compound
nucleus will not be reliable in this situation.
There also exist several analyses of experimental data, which reject the
binary decay mechanism of fragment production via sequential evaporation
from a compound nucleus at high excitation energy
\cite{hubele,Deses,napolit,ISIS}.
On the other hand, the picture of a nearly simultaneous
break-up in some freeze-out volume is justified in this case.
Indeed, the time scales of less than 100 fm/c are extracted for
multifragmentation reactions from experimental data
\cite{beaulieu,karnaukhov}.

\subsection{\bf Statistical Multifragmentation Model }
\label{subsec:SMM}
After completion of the cascade stage, when all produced particles leave the interaction zone, an excited RN is treated as thermalized. On the next stage such a thermalized many body system can emit multiple fragments that is described in the framework of the Statistical Multifragmentation Model (SMM) \cite{SMM-3}.The reason is that this
model was primary constructed for using after initial dynamical stage, and
adjusted for this kind of hybrid Monte-Carlo calculations.

The model assumes statistical equilibrium of excited nuclear system with
mass number $A_0$, charge $Z_0$, and excitation energy (above the ground
state) $E_0$ at a low-density freeze-out volume. This volume can be
parameterized as $V=V_0+V_f$, so the baryon density is
$\rho = A_0 / V$. $V_0$ is the volume of the system at the normal nuclear
density $\rho_0 \approx$ 0.15 fm$^{-3}$. $V_f$ is the so-called free
volume available for translational motion of fragments.
Note that the hypothesis of the statistical equilibrium, including the detail
balance principle, suggests that the short-range strong nuclear forces is not
responsible for the primary fragment formation beyond the freeze-out volume.
The model considers all break-up channels (ensemble of partitions $\{p\}$)
composed of nucleons and excited fragments taking into account the
conservation of baryon number, electric charge and energy.
An important advantage of the SMM is that besides these break-up
channels it includes also the compound nucleus channel, and takes into
account competition between all channels. In this way the SMM includes
the conventional evaporation and fission processes at low excitation
energy, and provides natural generalization of the de-excitation process
for high excitation energy.

In the model light nuclei with mass number $A \leq 4$ and charge
$Z \leq 2$ are treated as elementary stable particles with masses and
spins taken from the nuclear tables ("nuclear gas"). Only translational
degrees of freedom of these particles contribute to the entropy of the
system. Fragments  with $A > 4$ are treated as heated nuclear liquid
drops. In this way one may study the nuclear liquid-gas coexistence in
the freeze-out volume. Their
individual free energies $F_{AZ}$ are parameterized as a sum of the bulk,
surface, Coulomb and symmetry energy contributions
\begin{equation}
F_{AZ}=F^{B}_{AZ}+F^{S}_{AZ}+E^{C}_{AZ}+E^{sym}_{AZ}.
\end{equation}
The standard expressions for these terms are:
$F^{B}_{AZ}=(-W_0-T^2/\epsilon_0)A$, where $T$ is the temperature,
the parameter $\epsilon_0$ is related to the level density, and
$W_0 = 16$~MeV is the binding energy of infinite nuclear matter;
$F^{S}_{AZ}=B_0A^{2/3}(\frac{T^2_c-T^2}{T^2_c+T^2})^{5/4}$, where
$B_0=18$~MeV is the surface coefficient, and $T_c=18$~MeV is the critical
temperature of infinite nuclear matter; $E^{C}_{AZ}=cZ^2/A^{1/3}$, where
$c=(3/5)(e^2/r_0)(1-(\rho/\rho_0)^{1/3})$ is the Coulomb parameter, with the
charge unit $e$ and $r_0$=1.17 fm;  $E^{sym}_{AZ}=\gamma (A-2Z)^2/A$, where
$\gamma = 25$~MeV is the symmetry energy parameter.
These parameters are those of the Bethe-Weizs\"acker formula and correspond
to the assumption of isolated fragments with normal density in the
freeze-out configuration, an assumption found to be quite successful in
many applications. It is to be expected, however, that in a
more realistic treatment primary fragments will have to be considered
not only excited but also expanded and still subject to a residual nuclear
interaction between them.
These effects can be accounted for in the fragment
free energies by changing the corresponding liquid-drop parameters.
The Coulomb interaction of fragments in the freeze-out volume is
described within the Wigner-Seitz approximation (see ref. \cite{SMM-3}
for details).

As is well known, the number of partitions
of medium and heavy systems $(A_0\sim 100)$ is enormous
(see e.g. \cite{Jackson}). In order to take them into account the
model uses few prescriptions.
At small excitation energies the standard SMM code \cite{SMM-3} uses a
microcanonical treatment, however, taking into account a limited
number of disintegration channels: as a rule, only partitions with total
fragment multiplicity
$M \leq 3$ are considered. This is a very reasonable approximation at
low temperature, when the compound nucleus and low-multiplicity channels
dominate. Recently, a full microcanonical version of the SMM using
the Markov Chain method was introduced \cite{Jackson,Botvina01}. It can
be used for exploring all partitions without limitation.
However, it is a more time consuming approach, and it is used in special
cases only \cite{Botvina01}.

Within the microcanonical ensemble the
statistical weight of a partition $p$ is calculated as
\begin{eqnarray}
W_{\rm p} \propto exp~S_{\rm p},
\end{eqnarray}
where $S_{\rm p}$ is the corresponding entropy, which depends on fragments
in this partition, as well as on the excitation energy $E_0$, mass
number $A_{0}$, charge $Z_{0}$, volume $V$ of the system. In the standard
treatment we follow a description which corresponds to approximate
microcanonical ensemble. Namely, we introduce a temperature $T_{p}$
characterising all final states in each partition $p$. It is determined
from the energy balance equation taking into account the total excitation
energy $E_0$ \cite{SMM-3}. In the following we determine $S_{\rm p}$ for the
found $T_{p}$ by using conventional thermodynamical relations. In
the standard case, it can be written as
\begin{eqnarray}
S_{\rm p}=ln(\prod_{A,Z}g_{A,Z})+ln(\prod_{A,Z}A^{3/2})
-ln(A_0^{3/2})-ln(\prod_{A,Z}n_{A,Z}!)+
\nonumber
\\
(M-1)ln(V_f/\lambda_{T_{p}}^3)
+1.5(M-1)+\sum_{A,Z}(\frac{2T_{p}A}{\epsilon_0}-
\frac{\partial F^{S}_{AZ}(T_{p})}{\partial T_{p}}) , \nonumber
\end{eqnarray}
where $n_{A,Z}$ as the number of fragments with mass $A$ and charge
$Z$ in the partition,
$g_{A,Z}=(2s_{A,Z}+1)$ is the spin degeneracy factor,
$\lambda_{T_{p}}=\left(2\pi\hbar^2/m_NT_{p}\right)^{1/2}$ is the nucleon
thermal wavelength ($m_N\approx 939$ MeV is the average nucleon mass),
and the summation is performed over all fragments of the partition $p$.
We enumerate all considered partitions and select one of them according
to its statistical weight by the Monte-Carlo method.

At high excitation energy the standard SMM code makes a transition to
the grand-canonical ensemble \cite{SMM-3}, since the number of partitions
with high probability becomes too large.
In the grand canonical formulation, after integrating out translational
degrees of freedom, one can write the mean multiplicity of nuclear
fragments with $A$ and $Z$ as
\begin{eqnarray}
\label{naz} \langle n_{A,Z} \rangle =
g_{A,Z}\frac{V_{f}}{\lambda_T^3}A^{3/2} {\rm
exp}\left[-\frac{1}{T}\left(F_{AZ}(T,V)-\mu A-\nu Z\right)\right].
\end{eqnarray}
Here the temperature $T$ can be found from the total energy balance of the
system by taking into account all possible fragments with $A$ from 1
to $A_0$ and with $Z$ from 0 to $Z_0$ \cite{SMM-3}.
The chemical potentials $\mu$
and $\nu$ are found from the mass and charge constraints:
\begin{equation} \label{eq:ma2}
\sum_{A,Z}\langle n_{A,Z}\rangle A=A_{0},~~
\sum_{A,Z}\langle n_{A,Z}\rangle Z=Z_{0}. \nonumber
\end{equation}
In this case the grand canonical occupations $\langle n_{A,Z} \rangle$ are
used for Monte-Carlo sampling of the fragment partitions \cite{SMM-3}.
These two methods of partition generation are carefully adjusted to provide
a smooth transition from the low energy to the high energy regimes.

\subsection{\bf Propagation and de-excitation of hot fragments }

After the Monte-Carlo generation of a partition the temperature of the hot
fragments, their excitation energy and momenta can be found from the energy
and momentum balance. The Coulomb acceleration and propagation of fragments
must be taken into account also. In order to evaluate it the fragments are
placed randomly in the freeze-out volume $V$ (without overlapping), and
their positions are adjusted by taking into
account that their Coulomb interaction energy must be equal to the value
calculated in the Wigner-Seitz approximation. In the
following we resolve the Hamilton equations for motion of fragment from these
initial positions in their mutual Coulomb field.
The energy and momentum balances are strictly respected
during this dynamical propagation.

The secondary de-excitation of primary hot fragments includes
several mechanisms. For light primary  fragments  (with  $A\leq 16$)
produced in multifragmentation
even a  relatively  small  excitation  energy  may  be  comparable
with their total binding energy. In this case we  assume  that  the
principal mechanism of de-excitation is the explosive decay of the
excited nucleus into several smaller  clusters  (the  Fermi
break-up) \cite{fermi,Botvina87,SMM-3}.
In this decay the statistical weight of the channel $p$ containing
$n$  particles  with  masses $m_{i}$ ($i=1,\cdots,n$) in volume
$V_{p}$ can be calculated  in microcanonical approximation :
\begin{equation} \label{eq:Fer}
\Delta\Gamma_{p}\propto
\frac{S}{G}\left(\frac{V_{p}}{(2\pi\hbar)^{3}}\right)^{n-1}
\left(\frac{\prod_{i=1}^{n}m_{i}}{m_{0}}\right)^{3/2}\frac{(2\pi)^
{\frac{3}{2}(n-1)}}{\Gamma(\frac{3}{2}(n-1))}\cdot
\left(E_{kin}-U_{p}^{C}\right)^{\frac{3}{2}n-\frac{5}{2}},
\end{equation}
where $m_{0}=\sum_{i=1}^{n}m_{i}$ is the mass of the decaying nucleus,
$S=\prod_{i=1}^{n}(2s_{i}+1)$ is the degeneracy factor
($s_{i}$ is the $i$-th particle spin), $G=\prod_{j=1}^{k}n_{j}!$
is the particle identity factor ($n_{j}$ is the number of particles of
kind $j$). $E_{kin}$ is the total kinetic energy of  particles at
infinity which can be found through the energy balance by taking
into account the fragment excitation energy, $U_{p}^{C}$ is the Coulomb
barrier for this decay. We have slightly modified this model \cite{Botvina87}
by including fragment excited states stable with respect to the
nucleon emission as well as some long-lived unstable nuclei.

The successive particle emission from hot primary fragments with
$A>16$ is assumed to be their basic de-excitation mechanism, as in the
case of the compound nucleus decay.
Due to the high excitation energy of these fragments, the standard
Weisskopf evaporation scheme was modified to take into account the
heavier ejectiles up to $^{18}$O, besides light particles (nucleons,
$d$, $t$, $\alpha$), in ground and particle-stable excited states
\cite{Botvina87}. The width for the emission of a particle $j$ from
the  compound  nucleus $(A,Z)$ is given by:
\begin{equation} \label{eq:eva}
\Gamma_{j}=\sum_{i=1}^{n}\int_{0}^{E_{AZ}^{*}-B_{j}-\epsilon_{j}^{(i)}}
\frac{\mu_{j}g_{j}^{(i)}}{\pi^{2}\hbar^{3}}\sigma_{j}(E)
\frac{\rho_{A^{'}Z^{'}}(E_{AZ}^{*}-B_{j}-E)}{\rho_{AZ}(E_{AZ}^{*})}EdE.
\end{equation}
Here the sum is taken over the ground and all particle-stable excited states
$\epsilon_{j}^{(i)}~(i=0,1,\cdots,n)$ of the fragment $j$,
$g_{j}^{(i)}=(2s_{j}^{(i)}+1)$   is  the
spin degeneracy factor of the $i$-th excited  state,
$\mu_{j}$ and $B_{j}$ are corresponding reduced mass and separation energy,
$E_{AZ}^{*}$ is the excitation energy of the initial nucleus,
$E$ is the kinetic energy of an emitted particle in the centre-of-mass
frame. In eq. (\ref{eq:eva}) $\rho_{AZ}$ and $\rho_{A^{'}Z^{'}}$ are
the level densities of the initial $(A,Z)$ and final $(A^{'},Z^{'})$
compound  nuclei.  The cross section $\sigma_{j}(E)$ of the inverse
reaction $(A^{'},Z^{'})+j=(A,Z)$ was calculated using the optical model
with nucleus-nucleus potential \cite{Botvina87}. The evaporation
process was simulated by the Monte Carlo method and the conservation of
energy and momentum was strictly controlled in each emission step.

An important channel of de-excitation of heavy nuclei ($A>100$) is
fission. This process competes with particle emission, and it is also
simulated with the Monte-Carlo method.
Following the Bohr-Wheeler statistical approach we assume that
the partial width for the compound nucleus fission is proportional
to the level density at the saddle point $\rho_{sp}(E)$ \cite{SMM-3,Eren2013} :
\begin{equation} \label{eq:fis}
\Gamma_{f}=
\frac{1}{2\pi\rho_{AZ}(E_{AZ}^{*})}\int_{0}^{E_{AZ}^{*}-B_{f}}
\rho_{sp}(E_{AZ}^{*}-B_{f}-E)dE,
\end{equation}
where $B_{f}$ is the height of the fission barrier which is determined by
the Myers-Swiatecki prescription. For approximation of
$\rho_{sp}$ we used the results of the extensive analysis of nuclear
fissility and $\Gamma_{n}$/$\Gamma_{f}$ branching ratios. Concerning
masses, charges and energies of produced fission fragments see Refs.
\cite{SMM-3,Eren2013} for details.

All these models for secondary de-excitation
were tested by numerical comparisons with experimental data on
decay of compound nuclei with excitation energies less than 2--3 MeV
per nucleons. It is important that after all stages the SMM provides
event by event simulation of the whole break-up process and allows for
direct comparison with experimental events.

\subsection{\bf Experimental verification of SMM and prospects of the statistical approach }

As was shown already in first publications \cite{SMM-3,Botvina90}
the SMM gives very good description
of experimental data in the case when fragments are emitted from
equilibrated sources. Later on, many experimental groups have
successfully applied SMM for interpretation of their data.
There were convincing comparisons with experimental data in heavy ion
collisions around Fermi-energy \cite{MSU,INDRA,jandel,Iglio,Souliotis}.
\begin{figure}[ht]
\begin{center}
\begin{tabular}{cc}
\includegraphics[width=67mm]{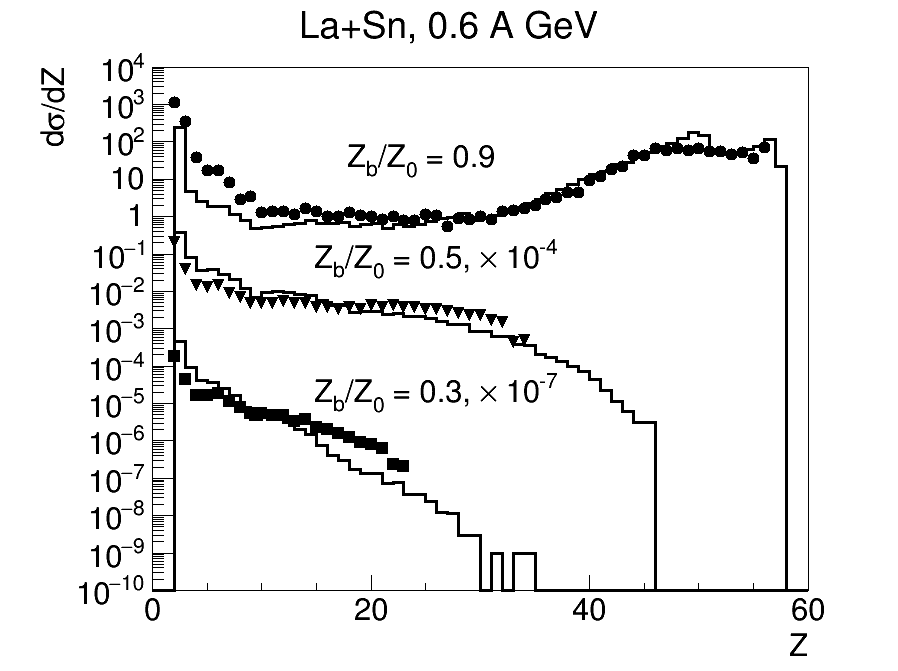}
&
\includegraphics[width=58mm]{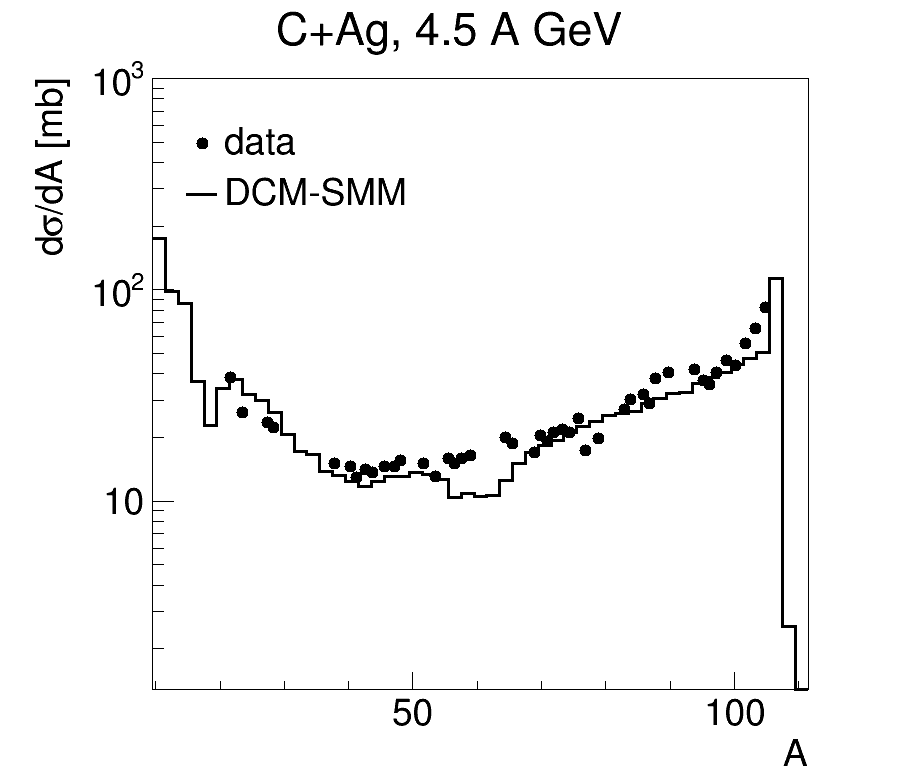}
\end{tabular}
\end{center}
\vspace{-5mm}
\caption{Left: Fragmentation of 0.6 AGeV projectile La nucleus. Data are from \cite{Ogul11}. Right: Fragment mass distribution in CAg collisions at 4.5 AGeV/c. Data are from \cite{Ahmad_CAg}}
\label{LaSn}
\end{figure}
In relativistic ion collisions the analyses have also demonstrated an
excellent performance of SMM for description of the nuclear residues
disintegration \cite{ALADIN,Xi97,Ogul11,EOS1,EOS2}. As well as for
the reaction initiated by light relativistic projectiles on heavy
nuclei \cite{ISIS,FAZA}.
It was demonstrated, that SMM describes charge (mass) distributions of
produced fragments and their evolution with excitation energy, isotope
distributions, multiplicities of produced particles and fragments in events,
charge distributions of first, second, third fragments in
the system, correlation functions (charge, angle, velocity ones) of
the fragments, fragment kinetic energy distributions. Simultaneously,
this model reproduces global characteristics of the systems, such as
caloric curves, critical indexes for the phase transition, different
moments of the fragment charge distribution.
\begin{figure}[hb]
\begin{center}
\begin{tabular}{cc}
\includegraphics[width=63mm]{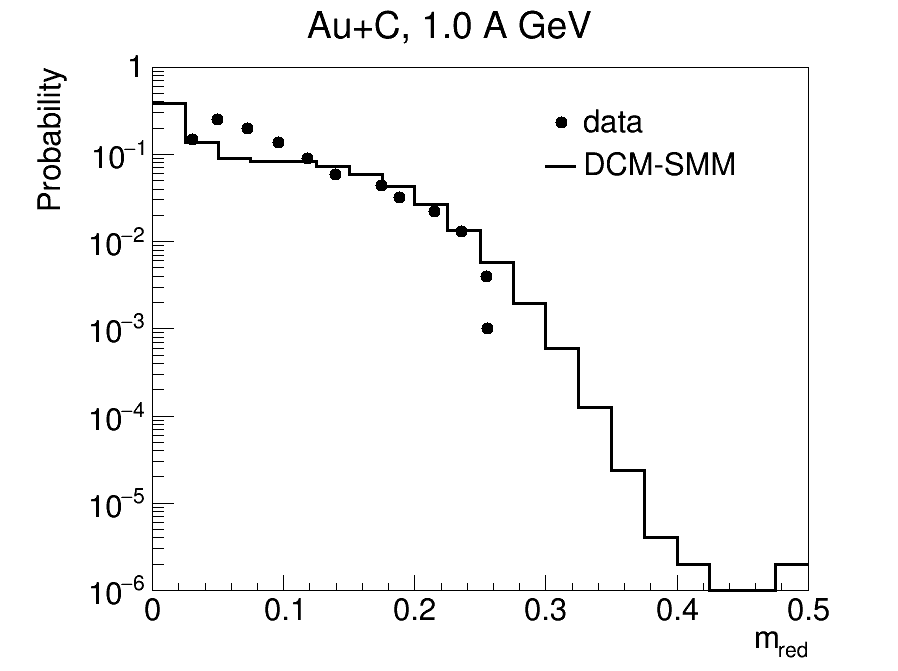}
&
\includegraphics[width=63mm]{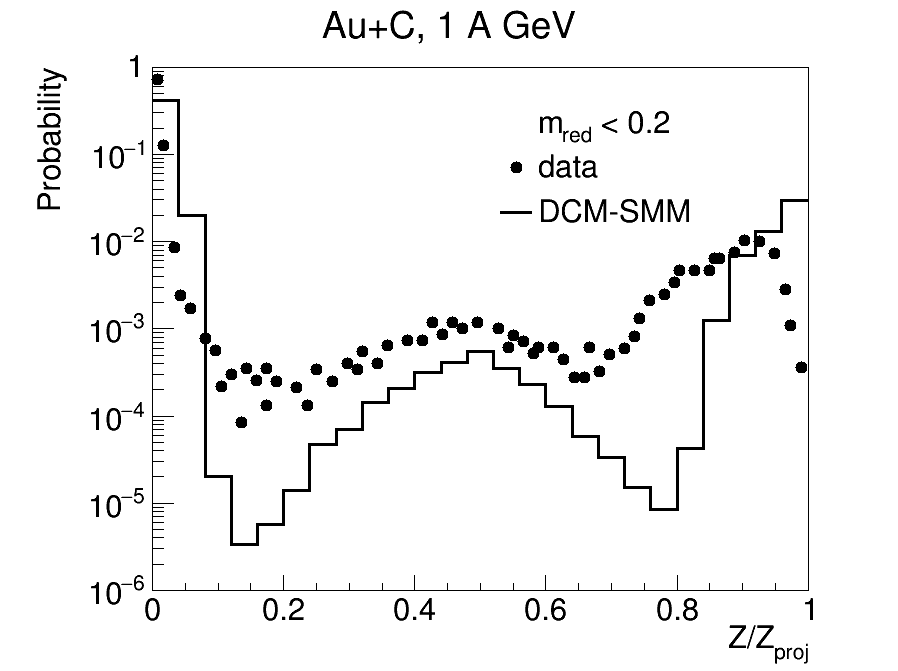}
\end{tabular}
\begin{tabular}{cc}
\includegraphics[width=63mm]{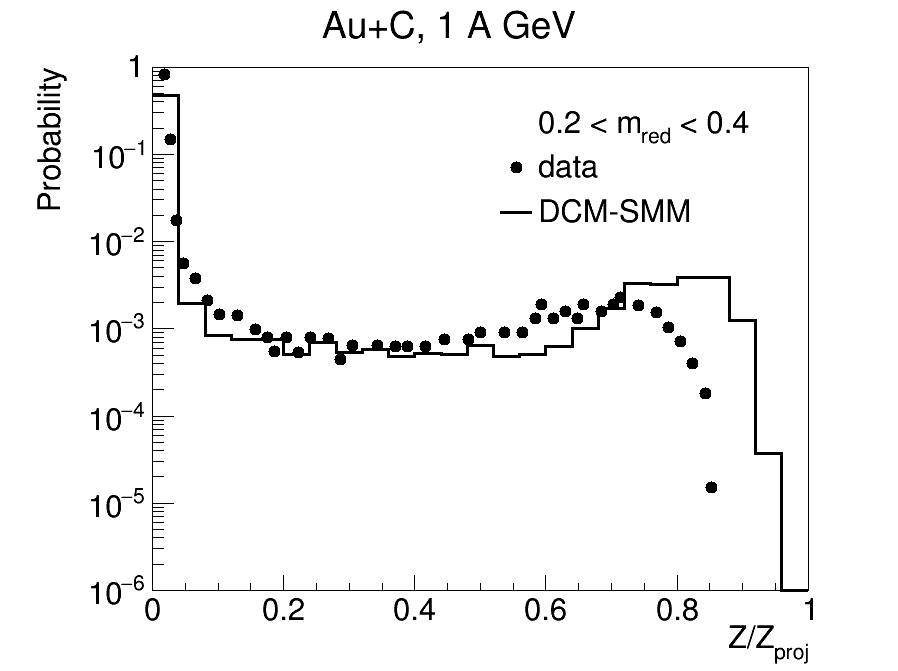}
&
\includegraphics[width=63mm]{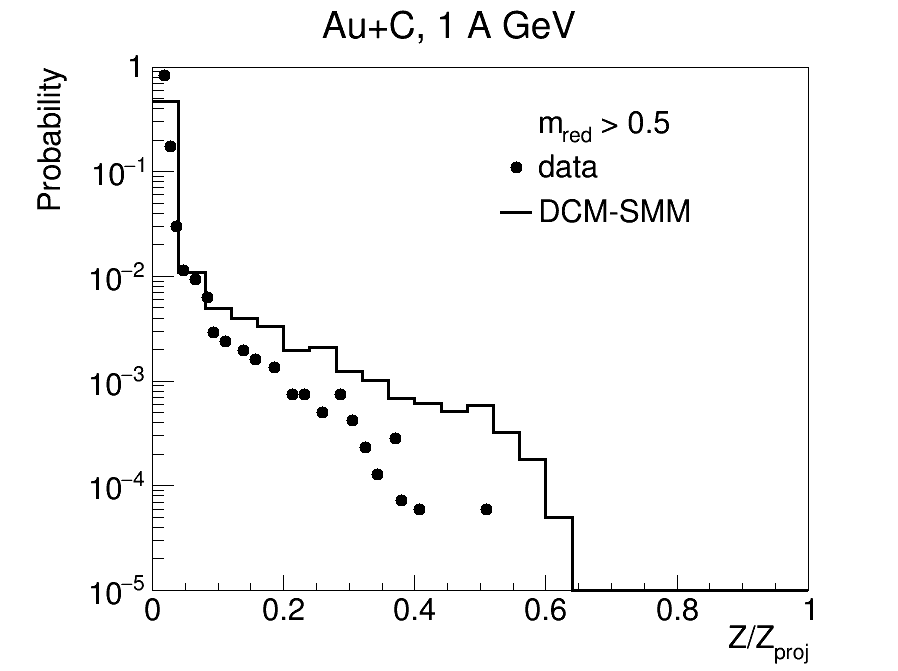}
\end{tabular}
\end{center}
\vspace{-5mm}
\caption{Fragmentation of projectile Au in AuC collisions at 1 AGeV. Top left: Fragment yield versus reduced charged fragment multiplicity, $m_{red}=m/Z_{proj}$. Other three plots are fragment charge distributions of Au as a function of $Z/Z_{proj}$ for three reduced multiplicity intervals. Data are from \cite{EOSd}.}
\label{AuC-1GeV}
\end{figure}
In other words, the model can describe almost completely experimental events of fragmentation and
multifragmentation. Charged fragments produced in heavy ion collisions in the energy range from hundreds MeV to few GeV in comarison with data are shown in Figures \ref{LaSn} and \ref{AuC-1GeV}. In Fig. \ref{LaSn} the cross sections $d\sigma/dZ$ for fragment production measured at ALADIN spectormeter was initiated by 600 AMeV  $^{124}$La projectiles directed onto reaction target consisting of $^{nat}$Sn \cite{Ogul11}. The data are sorted into two bins of the reduced bound charge $Z_{bound}$ = $Z\geq2$, where  $Z_{0}=Z_{proj}$.
The right plot in this Figure shows the mass number distribution in reaction CAg with 4.5 AGeV/c carbon beam \cite{Ahmad_CAg}. The multifragmentation of 1 AGeV Au incident on carbon together with data measured by EOS collaboration is demonstrated in Fig. \ref{AuC-1GeV}. The charged fragment multiplicity is shown in top left plot in the form of reduced multiplicity distribution, where $m_{red}=m/Z_{projectile}$. Three other plots represent the fragment charge distributions for three reduced multiplicity intervals. We must note that in the comparisons in Figures \ref{LaSn} and \ref{AuC-1GeV} with ALADIN and EOS data we did not take into account the experimental filters for particle at fragment detection, since it is unknown to us. For example, we expect that the extracted (in the calculations) intervals in $m_{red}$ in Fig.~\ref{AuC-1GeV} will be shifted to higher values after this correction.
 This effect and implementation of the trigger conditions will lead to the corresponding improvement of the fragment yields' comparison. More systematic comparison would require the collaboration with experimenters. However, the presented comparison shows that we correctly reproduce the main trends of fragment production within DCM+SMM. We have also found that using evaporation/fission processes only, GEM model, give us a qualitative disagreement with the data.

An important application of this statistical approach is related to
the production of hypermatter and hypernuclei from the excited residues.
These hyper-residues can be produced by the capture of strange particles
during the cascade stage of the relativistic collisions \cite{Botvina11,Botvina17}. The extension of SMM
into hypernuclear sector predicts the possibility to form many novel
hypernuclei, including exotic ones and multi-strange nuclei, which are not possible to produce in other reactions \cite{Botvina07,Buyuk13,Botvina16}.
There were realistic estimates of yields of the hypernuclei which can
help in preparation of experiments \cite{Botvina13}. Moreover, it was
demonstrated that by measuring the statistically produced hypernuclei
one can extract information about their properties, in particular, the
binding energy \cite{Buyuk18}. The mechanism of the hyperon capture will be implemented in the next version of DCM-SMM.

\section{\bf Dileptons production}
Dileptons are a unique tool to study the properties of hot and dense matter created in nuclear collisions. They might serve as probes for the in-medium properties of vector mesons and the predicted restoration of chiral symmetry. Unlike hadrons, the lepton pairs produced in the nuclear fireball do not participate in the strong interaction and therefore penetrate the strongly interacting medium with negligible final-state reactions. Thus we gain insight into all the different stages of a nuclear collision, from the first nucleon-nucleon interactions to the final freeze-out. But this also means that in experimental measurements we obtain time-integrated
spectra only, stemming from a broad variety of sources. In consequence we need good models that help to understand the production mechanisms and their contribution to the total spectra.  In particle-nucleus collisions transport models have been successful in describing the experimentally measured dilepton spectra. However, for a hot and dense environment as created in relativistic heavy-ion collisions it is supposed that medium effects play a crucial role for dilepton production. Measurements of emission of dielectrons  in nucleus-nucleus collisions at wide range of collision energy revealed an enhancement of invariant mass spectra of di-leptons yield in the interval 0.2 - 0.6 GeV. This enhancement was interpreted as in-medium modifications of hadronic resonances at high temperature and density resulting in strong broadening of the  $\rho$--meson and/or its ``mass--dropping''. These effects, in principle, can be implemented in transport models where all sources of dilepton emission are produced during the evolution of particle production in a nucleus-nucleus collision.

 As a first step, the analysis of di-electron production in heavy ion collisions  in the framework of the DCM-QGSM without any modifications of  was performed in \cite{gudtitov,nica}. In the current model (DCM-SMM) all important channels for the direct decay of vector mesons as well as for the Dalitz meson decays are considered. The direct decays of vector mesons $\rho \rightarrow e^{+}e^{-}$, $\omega \rightarrow e^{+}e^{-}$ and $\phi \rightarrow e^{+}e^{-}$ are taken into account. The main channels of the Dalitz decay of hadrons which contribute to the dilepton yield are $\pi^{0} \rightarrow \gamma e^{+}e^{-}$, $\eta \rightarrow \gamma e^{+}e^{-}$, $\omega \rightarrow \pi^{0} e^{+}e^{-}$ and $\Delta \rightarrow  Ne^{+}e^{-}$.
\begin{figure}
\begin{center}
\includegraphics[width=90mm]{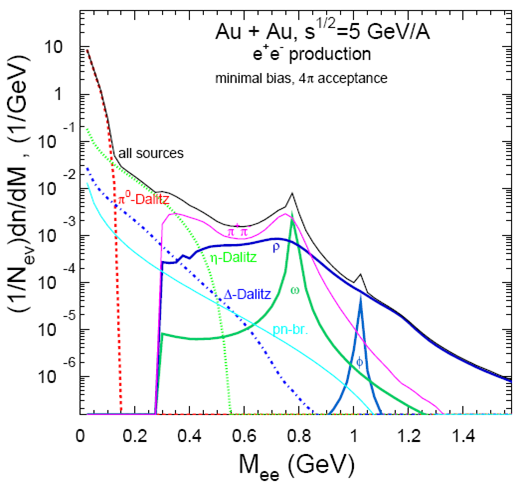}
\caption{Invariant mass distribution of $e^{+}e^{-}$ pairs in AuAu collisions at $\sqrt{s}=$ 5 GeV/nucleon .
}
\end{center}
\label{dilep-plot}
\end{figure}
 Additional sources of dilepton production are bremsstrahlung ($pn\rightarrow pne^{+}e^{-}$) and annihilation ($\pi^{+}\pi^{-}\rightarrow e^{+}e^{-}$) channels. An analysis of the time dependence of the dilepton creation rate for direct decay of vector mesons shows that the overwhelming part of dlleptons is emitted from the compressed region and, therefore, should suffer some medium effect. Fig. 6 demonstrates invariant mass distribution of $e^{+}e^{-}$ pairs coming from different sources in AuAu collisions with effect of widening of the width of $\rho$ and $\omega$ resonances in the model. We plan to develop the model taking into account the modification of hadron properties in more details.


\section{\bf Lambda polarization and vorticity}
\label{sec-1}

In non-central relativistic heavy ion collisions, a strong vorticity field is generated in the produced matter as a result of the large orbital angular momentum that is brought into the system.  This vorticity field can lead to the polarization of particles of non-zero spin along the direction of the vorticity field due to their spin-orbit or spin-vorticity coupling. Measurements of the global spin polarization of $\Lambda$ hyperons by the STAR Collaboration~\cite{STAR:2017ckg,PhysRevC.98.014910} have confirmed the existence of the most vortical fluid ever known, with an average vorticity of more than $10^{21}$ s$^{-1}$.

There are several definitions of the vorticity used in the literature that are suitable
for analyzing different aspects of the rotation effects. In the present study we consider two of them~\cite{Baznat:2013zx,Baznat:2015eca,Teryaev:2015gxa}. The first one is the relativistic kinematic vorticity
   \begin{eqnarray}
   \label{rel.kin.vort.}
   \omega_{\mu\nu} = \frac{1}{2}
   (\partial_{\nu} u_{\mu} - \partial_{\mu} u_{\nu}),
   \end{eqnarray}
where $u_{\mu}$ is a collective local four-velocity of the matter
and the second one is so-called thermal vorticity
   \begin{eqnarray}
   \label{therm.vort.}
   \varpi_{\mu\nu} = \frac{1}{2}
   (\partial_{\nu} \hat{\beta}_{\mu} - \partial_{\mu} \hat{\beta}_{\nu}),
   \end{eqnarray}
where $\hat{\beta}_{\mu}=\hbar\beta_{\mu}$ and $\beta_{\mu}=u_{\nu}/T$
with $T$ being the local temperature and $\varpi$ is dimensionless.

These two methods are used to calculate the Lambda polarization of hyperons.
The first one is anomalous mechanism of hyperon polarization related to kinematical vorticity and helicity.
The polarization is related \cite{Baznat:2013zx,Sorin:2016smp}  to the strange axial charge
\begin{equation}
\label{q5}
Q_5^s=N_c \int d^3 x c_v \gamma^2 \epsilon^{i j k}v_{i}
\partial_{j}v_ k.
\end{equation}
$c_v$ is the chiral vorticity coefficient describing the axial
vortical effect
\begin{equation}
\label{c_v}
c_v = \frac{\mu^{2}_{s}}{2\pi^{2}} + k\frac{T^{2}}{6},
\end{equation}
where the second term is temperature-dependent with adjustable parameter $k$.
As a result the quark and hadronic observables are related, that is of
special importance in the confined phase.
For polarization we get the formula
\begin{equation}
\label{ratio} <\Pi_0^{\Lambda}>\, = \, \frac{m_\Lambda \,
\Pi_0^{\Lambda,lab}}{p_y}\,= \, <\frac{m_{\Lambda}}{N_{\Lambda}
\,p_y }> Q_5^s \,\equiv \, <\frac{m_{\Lambda}}{N_{\Lambda} \,p_y }>
\frac{N_c}{{2 \pi^2}} \int d^3 x \,\mu_s^2(x) \gamma^2 \epsilon^{i j
k}v_{i} \partial_{j}v_ k.
\end{equation}

In local thermal equilibrium, the
ensemble average of the spin vector for spin-$1/2$ fermions with
four-momentum $p$ at space-time point $x$ is obtained from the statistical-hydrodynamical
model \cite{Becattini:2013fla} as well as the Wigner function approach
\cite{Fang:2016vpj} and reads
\begin{equation}
S^{\mu}(x,p)=-\frac{1}{8m}\left(1-n_{F}\right)\epsilon^{\mu\nu\rho\sigma}p_{\nu}\varpi_{\rho\sigma}(x),\label{eq:spin_thermal}
\end{equation}
where $\varpi_{\mu\nu}$ is the thermal vorticity
and $\beta^{\mu}=u^{\mu}/T$ being the inverse-temperature four-velocity.
In Eq. (\ref{eq:spin_thermal}), $m$ is the mass of the particle
and $n_{F}=1/[1+\exp(\beta\cdot p\mp\mu/T)]$ is the Fermi-Dirac distribution
function for particles ($-$) and anti-particles ($+$).

The spin vector $S^{\mu}(x,p)$  is defined in the
center of mass (CM) frame of Au+Au collisions. In the STAR experiment,
the $\Lambda$ polarization is measured in the local rest frame of
the $\Lambda$ by its decay proton's momentum. The spin vector of
$\Lambda$ in its rest frame is denoted as $S^{*\mu}=(0,\mathbf{S}^{*})$
and is related to the same quantity in the CM frame by a Lorentz boost
\begin{equation}
\mathbf{S}^{*}(x,p)=\mathbf{S}-\frac{\mathbf{p}\cdot\mathbf{S}}{E_{p}\left(m+E_{p}\right)}\mathbf{p}.\label{eq:spin_rest_frame}
\end{equation}
By taking the average of $\mathbf{S}^{*}$ over all $\Lambda$ particles
produced at the freeze-out stage in the hydrodynamic picture of heavy ion collisions, we obtain the average spin
vector
\begin{equation}
\left\langle \mathbf{S}^{*}\right\rangle =\frac{1}{N}\sum_{i=1}^{N}\mathbf{S}^{*}(x_{i},p_{i}),
\end{equation}
where $N$ is the number of $\Lambda$s in all events and $i$ labels
one individual $\Lambda$.
\begin{wrapfigure}{r}{0.5\textwidth}
\includegraphics[width=60mm]{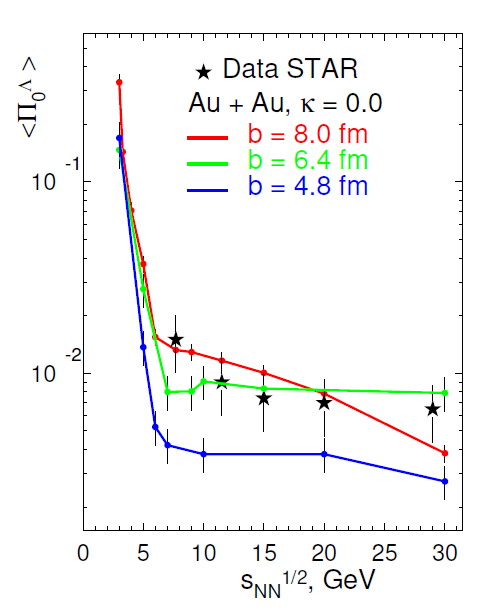}
\caption{Energy dependence of Lambda polarization in peripheral AuAu collisions for three values of impact parameter.
Experimental data are from STAR~\cite{STAR_Lam-pol}.
}
\label{Lambda-polar}
\vspace{-5mm}
\end{wrapfigure}
The global $\Lambda$ polarization in the
STAR experiment is the projection of $\left\langle \mathbf{S}^{*}\right\rangle $
onto the direction of global angular momentum in off-central collisions
(normal to the reaction plane),
\begin{equation}
P=2\frac{\left\langle \mathbf{S}^{*}\right\rangle \cdot\mathbf{J}}{|\mathbf{J}|},\label{eq:polarization}
\end{equation}
where we have included a normalization factor ($P$ is normalized
to 1) and $\mathbf{J}$ denotes the global orbital angular momentum
of off-central collisions.
In our calculations, some
relations between kinetic and hydrodynamic description were considered. Calculations include spatial and temporal dependence of the strange chemical potential. In numerical simulations the space-time is decomposed on cells allowing to define velocity and vorticity in the model. To define the strange chemical potential (assuming that $\Lambda$ polarization is carried
by strange quark) we used the matching procedure of distribution functions to its (local) equilibrium values. We also determined in this way the values of temperature. Energy dependence of $\Lambda$ polarization for three values of impact parameter together with STAR data is shown on Fig.~\ref{Lambda-polar}.

\section{\bf Numerical simulation and comparison to experimental data}
Results of simulations performed by the model is compared to the available experimental data  and the calculations using UrQMD-3.4 at the NICA energy range. Before comparison we outline similarity and differences between the models, DCM-SSM and UrQMD-3.4 \cite{urqmd-3.4}.
\subsection{\bf DCM-QGSM and URQMD: similarity and difference ~}
As the first (fast) stage of a collision in the DCM-SMM is simulated by the DCM and QGSM we compare the DCM-QGSM to UrQMD.
Both models are formulated as Monte-Carlo event generators
allowing to perform a careful analysis of the measurable quantities
by introducing  all necessary experimental cuts. Both treat the production of new particles via formation and fragmentation of specific colored objects, strings. Strings are uniformly stretched, with constant string tension $\kappa \approx 1$\,GeV/fm,  between the quarks, diquarks and their antistates.
To describe hadron-nucleus $(hA)$ and nucleus-nucleus $(A+A)$
collisions the momenta and positions of nucleons in the nuclei are
generated according to the Fermi momentum distribution and the
Wood-Saxon density distribution, respectively.
The propagation of particles is governed by Hamilton equation of
motion, and both models use the concept of hadronic cascade for the
description of $hA$ and $A+A$ interactions.

The differences between the models arise on different stages of a string formation and fragmentation. The UrQMD
belongs to the group of models based on classical FRITIOF model
\cite{fritiof}, while the DCM-QGSM uses the Gribov Reggeon field theory
(RFT) \cite{grt,PaSh07} that results in differences on the string formation step. The second stage concerns string fragmentation. The fragmentation functions which determine the energy, momentum, and the type of the hadrons produced during the string decay, are different in the models.
The third type of differences deals with the number and type of the stings produced
in the collision. Due to the different mechanisms of string excitation
and fragmentation, these numbers are also different for two
microscopic models. Last but not least, the models do not use the same tables of hadrons. The last versions of UrQMD (starting with version 2.3) were modified by including a continuous spectrum of high resonance states that results in the improved description of transverse momentum spectra of particles in heavy ion collision. The UrQMD contains 55 baryon and 32 meson states together with their antistates, whereas the QGSM takes into account octet and decuplet baryons, and nonets of vector
and pseudoscalar mesons, as well as their antiparticles.
Detailed comparison of the models DCM-QGSM and UrQMD was done in the article \cite{bravinac78}. We should note that essential shortcoming of the UrQMD, in comparison with the DCM-QGSM, is absence of the mechanism forming a residual nucleus and its subsequent disintegration.

In the following subsections the results of DCM-SMM are compared to data measured in nucleus-nucleus collisions at NICA energy range. We have concentrated on bulk observables like multiplicities, particle spectra to demonstrate the relevance of the model. Particle yields and their spectra in nucleus-nucleus collisions are compared with UrQMD calculations, as well.

\subsection{\bf Nucleus - nucleus collisions ~}
We focus on comparison of the model calculations to data measured in experiments performed at AGS and by collaboration NA49 at SPS which cover NICA energy range.
\subsubsection{\bf Particle yield ~}
Fig.~\ref{multip} shows the excitation function of mean multiplicities for different particle species in central AuAu/PbPb collisions.The model overestimates the yield of pions at the whole energy range and underestimates kaon multiplicities at $\sqrt{s}= 5-10$ GeV/n. In this energy range the enhanced yield of  $K^{+}-$mesons and hyperons were observed by collaboration NA49. This enhancement not described by transport models can indicate specific modification of hadron properties and their interactions inside a dense/hot nuclear matter.
\subsubsection{\bf Rapidity spectra ~}
We start  with the rapidity spectra of net protons emitted in AuAu/PbPb collisions measured at AGS and SPS energies. These spectra reflect
the baryon stopping that determines a part of the incident
energy of colliding nuclei deposited into a produced fireball
and hence into the production of secondary particles. Obviously, the number of collision per baryon increases with the mass number of the colliding nuclei, and hence the heaviest systems, such as Pb+Pb or Au+Au, are best suited for the creation of strongly stopped matter and high energy densities. A proper reproduction of the baryon stopping is extremely  important for theoretical understanding of the dynamics of the nuclear collisions. Figures.~\ref{AGS_prot_y-dist} and~\ref{NA49_prot_y-dist}  show calculated proton rapidity distributions in central AuAu/PbPb collisions compared to data from  from AGS to SPS at the same values of centralities. Distributions calculated by both DCM-SMM and UrQMD at 2, 4, 8 and 10.5 AGeV (AGS), and 20, 30, 40, 80 AGeV (NA49)  demonstrate more pronounced picks at mid-rapidity than the data. This enhancement  can be interpreted as an energy transfer overestimation by the model in central heavy ion collisions at AGS and at lower SPS energies. By this reason the deuteron spectra formed on the coalescence stage replicate the shape of proton spectra (Fig.~\ref{d-rap}).
There are additional picks on distributions given by UrQMD at projectile and target rapidities (positive/negative) since a residual nucleus is not formed after a collision stage in the model.

Since the bulk of produced particles are pions, their rapidity spectra depend on the energy deposited by nucleons in the course of collisions. Obviously, they are overestimated at mid-rapidity, as well, at the all AGS - SPS energy range (Figs.~\ref{pi-rap_AGS} and \ref{pi-rap_NA49}). This discrepancy, together with proton spectra, tells us that a hadronic transport model based on superposition of binary hadron-hadron collisions deviates from relevant description of central heavy ion collisions.

\subsubsection{\bf Transverse mass spectra~}
Figure \ref{mt-dist} showes the transfer mass spectra of $\pi^{+}$ and $K^{+}$ in central Pb+Pb collisions compared with UrQMD and NA49 data. The spectra given by DCM-SMM are softer than those from UrQMD and NA49 data. The sources of the discrepancy are the following. Given the hadron properties simulated by the quark-gluon string model are adjusted by comparison with experiments in pp-collisions this discrepancy, again, comes from modification of hadron properties and interactions in hot and dense nuclear matter. Agreement of the UrQMD with NA49 data was attained by extension of the model in version 2.3. To reproduce the experimentally measured high transverse mass values a modified treatment of a string decay with high mass resonances  was introduced \cite{urqmd-2.3}. From our point of view such extension must be justified by a physical mechanism resulting in such hardening of transverse spectra.
\section{\bf Discussion and summary}
We presented the new Monte-Carlo generator which is a combination of the cascade part of the DCM-QGSM and the Statistical Multifragentation Model, SMM. The SMM allows for natural extension of cascade-evaporation calculations for the fast multifragmentation processes. Its main assumption is that nuclear fragments are produced simultaneously in the explosive break-up of a thermalized nuclear system formed at the intermediate stage of a highly-dissipative nuclear reaction. It replaces preequilibrium and sequential evaporation parts of DCM-QGSM, which failure to describe intermediate mass fragment (IMF) production.

This combined model DCM-SMM is applied for simulation of heavy ion collisions at NICA energy range and compared with the data measured by the experiments at AGS and collaboration NA49. It satisfactory reproduces bulk properties of produced hadrons and nuclear fragments in heavy ion collisions and could serve as a good instrument in the stage of preparation of a new experiment and preliminary analysis of measurements. Further development of the model is connected with i) the improvement of description of transverse mass distributions by including heavier mass resonances, baryonic and mesonic; ii) taking into account the dependence of coalescence parameters from rapidity of coalesced baryons; iii) modification of hadron features on nuclear density.
\subsection*{\bf Acknowledgments. ~}
The authors pay a tribute to one of the founders and a developer of the model  prof. Konstantin Gudima who passed away in 2018. M.Baznat and G.Musulmanbekov thank O. Teryaev and O. Rogachevsky for stimulating discussions. A. Botvina acknowledges the support of BMBF (Germany). The work has been performed in the framework of the project 18-02-40084/19 supported by RFBR grant ``Megascience NICA''.
\appendix
\section{\bf Running the code}
The DCM-SMM program is available as an executable binary file
on UNIX/Linux platforms. A bash shell script file is
provided to define the input parameters and run the program (see A.1.)
The input parameters include number of jobs to run, number of
events per job, projectile and target charges and atomic numbers,
reference system (laboratory or equal velocity) and collision
energy, impact parameter range. As a result of simulation two
output files are created: *.inf and *.out, where ``*'' stands for
the output file name. The first one contains information about the
input parameters as well as some additional information
about the reaction, for example, geometric and inelastic cross
sections, the number of projectile and target participants, and the parameters used in the simulation. The second
file contains the characteristics of particles and nuclear fragments produced
on event-by-event basis (see A.2.).
Produced particles are identified by their lepton (LN), charge (EN), strange (SN) and baryonic (BN) numbers. Furthermore, they are assigned PDG identification codes, which are given in Table 1.
 \begin{table} [hb]
\caption{Particle Data Group (PDG) Monte Carlo particle identification numbers(corresponding antiparticles have negative sign)}
\begin{center}
\begin{tabularx}{\textwidth}{|
p{\dimexpr 0.17\linewidth-2\tabcolsep-2\arrayrulewidth}|
p{\dimexpr 0.17\linewidth-2\tabcolsep-\arrayrulewidth}|
p{\dimexpr 0.16\linewidth-2\tabcolsep-\arrayrulewidth}|
p{\dimexpr 0.16\linewidth-2\tabcolsep-\arrayrulewidth}|
p{\dimexpr 0.17\linewidth-2\tabcolsep-\arrayrulewidth}|
p{\dimexpr 0.17\linewidth-2\tabcolsep-\arrayrulewidth}|}
\hline
\textit{Particle} & \textit{PDG ID} & \textit{Particle} & \textit{PDG ID} & \textit{Particle} & \textit{PDG ID} \\
\hline
${\gamma}$ & 22 & K$^{0}$$_{\mathrm{L}}$ & 130 & ${\Sigma}$$^{+}$ & 3222 \\
\hline
\textit{e}$^{-}$ & 11 & K$^{0}$$_{\mathrm{S}}$ & 310 & ${\Sigma}$$^{0}$ & 3212 \\
\hline
${\nu}$$_{\mathrm{e}}$ & 12 & K$^{0}$ & 311 & ${\Sigma}$$^{-}$ & 3112 \\
\hline
${\mathrm{\mu}}$$^{{-}}$ & 13 & K$^{+}$ & 321 & ${\Sigma}$$^{\mathrm{*+}}$ & 3224 \\
\hline
${\nu}$$_{\mathrm{{\mathrm{\mu}}}}$ & 14 & K$^{-}$ & -321 & ${\Sigma}$$^{\mathrm{*0}}$ & 3214 \\
\hline
${\pi}$$^{0}$ & 111 & K$^{\mathrm{*0}}$ & 313 & ${\Sigma}$$^{\mathrm{*-}}$ & 3114 \\
\hline
${\pi}$$^{+}$ & 211 & K$^{\mathrm{*+}}$ & 323 & ${\Xi}$$^{0}$ & 3322 \\
\hline
${\pi}$$^{-}$ & -211 & p & 2212 & ${\Xi}$$^{-}$ & 3312 \\
\hline
${\rho}$$^{0}$ & 113 & n & 2112 & ${\Xi}$$^{\mathrm{*0}}$ & 3324 \\
\hline
${\rho}$$^{+}$ & 213 & ${\Delta}$$^{++}$ & 2224 & ${\Xi}$$^{\mathrm{*-}}$ & 3314 \\
\hline
${\rho}$$^{-}$ & -213 & ${\Delta}$$^{+}$ & 2214 & ${\Omega}$$^{{-}}$ & 3334 \\
\hline
${\eta}$ & 221 & ${\Delta}$$^{0}$ & 2114 &  &  \\
\hline
${\eta}$$^{\prime}$ & 331 & ${\Delta}$$^{-}$ & 1114 &  &  \\
\hline
${\omega}$ & 223 & ${\Lambda}$ & 3122 &  &  \\
\hline
${\varphi}$ & 333 &  &  &  &  \\
\hline
\end{tabularx}
\end{center}
\end{table}
Nuclear codes are given as 10-digit numbers ${\pm}$10LZZZAAAI. For a (hyper)nucleus consisting
 of n$_{\mathrm{p}}$ protons, n$_{\mathrm{n}}$ neutrons and n$_{\mathrm{{\Lambda}}}$ ${\Lambda}$’s,
 A = n$_{\mathrm{p}}$ + n$_{\mathrm{n}}$ + n$_{\mathrm{{\Lambda}}}$ gives the total baryon number,
 Z = n$_{\mathrm{p}}$ the total charge and L = n$_{\mathrm{{\Lambda}}}$ the total number of strange
 quarks. I gives the isomer level, with I = 0 corresponding to the ground state and I {\textgreater}
 0 to excitations, see [4], where states denoted m, n, p, q translate to I = 1 ${-}$ 4. As examples,
 the deuteron is 1000010020 and $^{235}$U is 1000922350 [16].
\subsection{\bf Input file \\}
In order to run the simulation user writes the input
parameters in the provided bash shell script file between lines "Begin
Input parameters" and "End Input parameters". The input parameters include
\begin{itemize}
  \item name of output files,
  \item name of executable file,
  \item number of jobs to run,
  \item number of events per job,
  \item projectile and target charges and atomic numbers,
  \item reference system (laboratory or equal velocity),
  \item collision energy,
  \item impact parameter interval.
\end{itemize}
An example of user editable part of the script is given below. The script creates a directory with a name defined by a variable "basename" and generates intermediate input files for running the program within it.

\begin{verbatim}
# The basename is the name of the folder for the output files which will
# be created by this script in the directory the script is called.
# The basename will also be in front of every outpufile to easily recognize it
#
# BEGIN Input parameters
basename='AuAu_ss9_mb'
exename='dcm_smm.exe'
jobs_per_energy=1
events_per_job=1000
#
AP="197."	# Projectile mass
AT="197."	# Target mass
ZP="79."	# Projectile charge
ZT="79."	# Target charge
BMIN="0.0"	# Minimum of impact parameter (fraction, 0 to 1)
BMAX="1.0"	# Maximum of impact parameter (fraction, 0 to 1)
KSYS=2		# Observer system (1 - lab sys, 2 -nucleon-nucleon cms)
E0="9.0"	# Energy (GeV): KSYS=1 -> E0=E_lab; KSYS=2 -> E0=sqrt(s)
#####
# END Input parameters
#
# Here the random seed is initialized
seed="date +%s"

INPUTFILE=$basename
touch $INPUTFILE
read -d '' str3 <<- EOF
$basename.inf
$basename.out
$AP, $AT, $ZP, $ZT, 0.0, 0.940, $E_coll, $N_events
$STAT
$BMIN, $BMAX, 1, $KSYS
#**************************************
EOF
echo  "$str3" > $INPUTFILE
\end{verbatim}

\subsection{\bf Output file *.out for a single event. \\}
The output file *.out begins with a header giving information about
the simulated collisions and brief description of the event structure
followed by lists of particles generated in each event. The event
header is a line containing an event number, number of particles after
cascade and coalescence part of the simulation, impact parameter and
its x and y components. The next line contains information about target
residual nucleus: number of fragments it decayed on, atomic number,
charge, strangeness, exitation energy and momentum components. Only the
number of fragments dould be used for further processing, the rest is for
ingormation only. The next lines in a number corresponding to that of the
fragments are describing the respective fragments: charge, lepton number,
strangeness, barion number, PDG ID, $p_{x}$, $p_{y}$, $p_{zcm}$, $p_{zlab}$,
and mass. These lines are followed by the same information about the projectile fragments and particles produced after cascade and coalescence stages of a reaction.

\begingroup
\footnotesize
\begin{verbatim}

Results of DCM-SMM calculations of nuclear  collisions
 of A1=197.,Z1= 79. + A2=197.,Z2= 79.
 at T0=  11.434(sqrt(s)=   5.003) GeV/nucleon in the collider
 (equal velocities=cms for A1=A2) system

Characteristics of event:
  No. of event, number of produced particles after cascade
  and light clusters after coalescence stages,  b,bx,by - impact parameter(fm)

  Target residual nucleus:
  Number of fragments (it decays on), its atomic number,
  charge, strangeness, excit. energy and 3-momentum

  Characteristics of fragments:
  charge, lepton number, strangeness, baryon number, PDGID,
  P(x), P(y), P(z), Plab(z), mass

  Projectile residual nucleus: the same as for target residual

  Characteristics of produced particles after cascade and light
  clusters after coalescence stages: the same as for fragments

        1         5 14.194 13.709  3.681
     5  194. 78. -0. 0.0154   0.1374   0.2630 450.4103
   0   0   0   1        2112   1.3901E-02 3.3045E-02 2.3014E+00 1.2286E+01 9.40000E-01
   0   0   0   1        2112  -9.4771E-03 4.2047E-02 2.1817E+00 1.1695E+01 9.40000E-01
   0   0   0   1        2112   5.3359E-02 3.1486E-02 2.2510E+00 1.2038E+01 9.40000E-01
   0   0   0   1        2112  -1.4019E-03-1.0408E-02 2.5417E+00 1.3481E+01 9.40000E-01
  78   0   0 190  1000781900   8.1056E-02 1.6684E-01 4.4116E+02 2.3536E+03 1.78600E+02
     5  195. 79. -0. 0.0230   0.0102   0.0697-452.0061
   0   0   0   1        2112   1.0937E-02 4.0639E-02-2.3661E+00-1.4645E-02 9.40000E-01
   0   0   0   1        2112  -1.6699E-02 3.9166E-02-2.2739E+00 2.0027E-02 9.40000E-01
   0   0   0   1        2112  -1.1652E-02-3.2099E-02-2.3680E+00-1.5650E-02 9.40000E-01
   0   0   0   1        2112  -6.1822E-03-1.3715E-02-2.2678E+00 2.1564E-02 9.40000E-01
  79   0   0 191  1000791910   3.3818E-02 3.5689E-02-4.4274E+02 4.5704E-01 1.79540E+02
   1   0   0   1        2212   2.9709E-01-3.6733E-01-2.1463E+00-2.1463E+00 9.38280E-01
   1   0   0   2  1000010020   1.4571E-01 4.1871E-01 4.6205E+00 4.6205E+00 1.87612E+00
   0   0   0   1        2112  -6.6378E-02-1.0530E-01 1.6096E+00 1.6096E+00 9.39570E-01
   0   0   0   1        2112  -3.6069E-01-4.2789E-01-2.2049E+00-2.2049E+00 9.39570E-01
  -1   0   0   0        -211  -1.6605E-01 1.7860E-01-1.2341E-01-1.2341E-01 1.39570E-01
\end{verbatim}
\endgroup


\begin{figure}[ht]
\begin{center}
\begin{tabular}{cc}
\includegraphics[width=60mm]{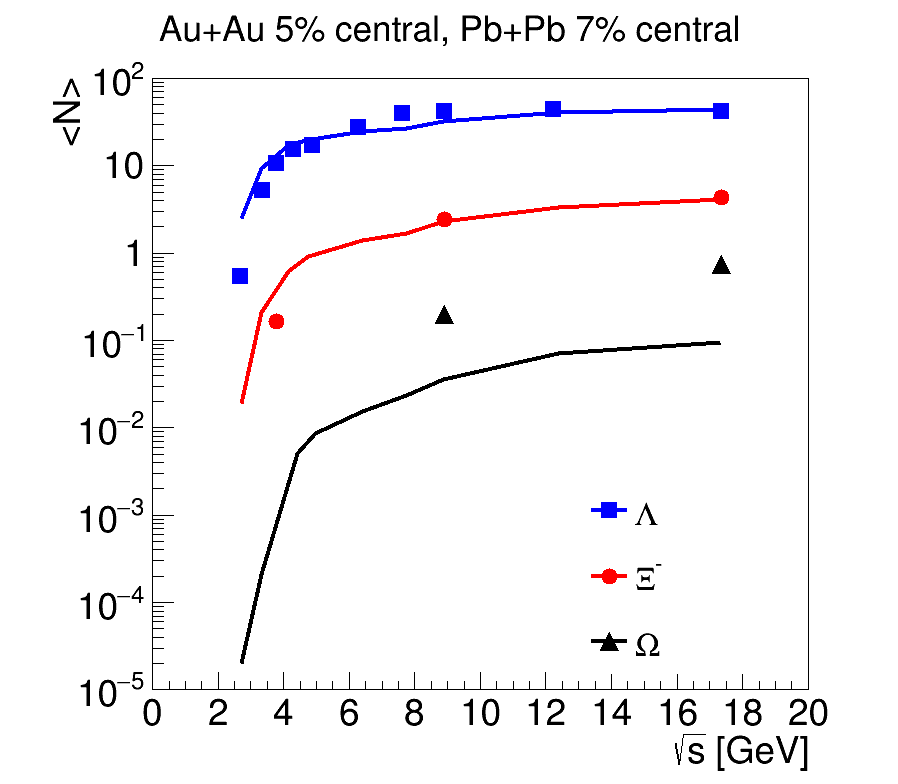}
&
\includegraphics[width=60mm]{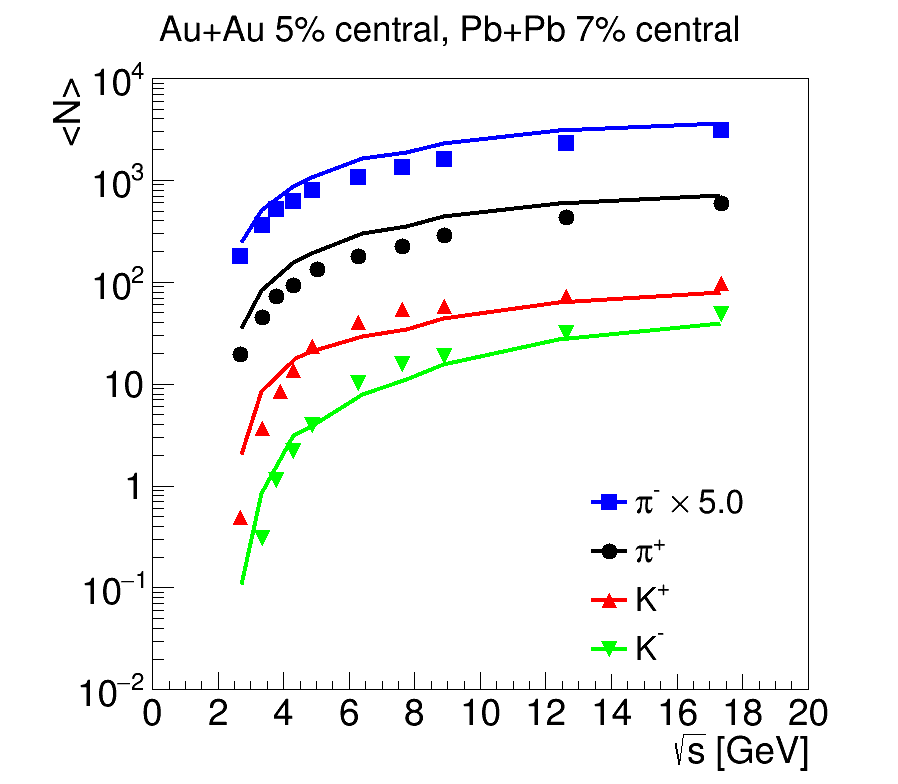}
\end{tabular}
\vspace{-3mm}
\caption{Excitation function of particle multiplicities in Au+Au/Pb+Pb collisions from Elab = 2 AGeV to 160 AGeV. Full lines are DCM-SMM calculations. The corresponding data from experiments \cite{E0895, E895.2, E895.3, na49.6, na49.7, na49.8, na49.9, na49.10, na49.11,na49.12, na49.13} are depicted with symbols.}
\end{center}
\labelf{multip}
\vspace{-5mm}
\end{figure}

\begin{figure}[ht]
\begin{center}
\includegraphics[width=110mm]{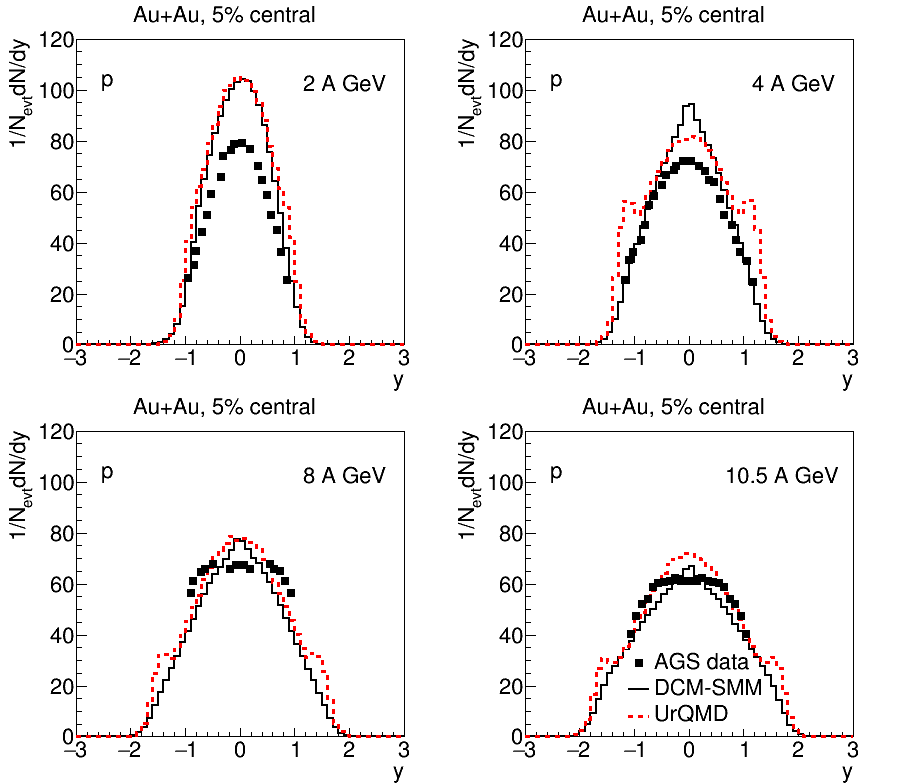}
\vspace{-3mm}
\caption{Rapidity spectra of protons for AGS energies from central collisions of Au+Au (AGS). Experimental data
are from the~\cite{E0895},~\cite{E877},~\cite{E917},~\cite{E866}. Black and red histograms are DCM-SMM and UrQMD calculations, correspondingly.
}
\end{center}
\labelf{AGS_prot_y-dist}
\vspace{-5mm}
\end{figure}
\begin{figure}[ht]
\begin{center}
\includegraphics[width=110mm]{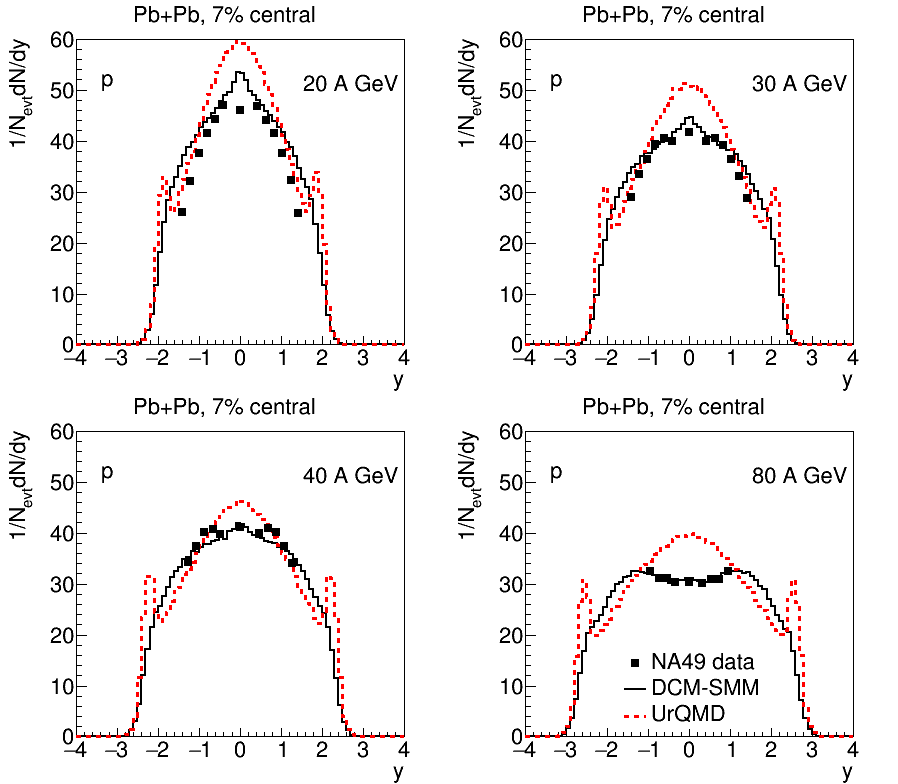}
\vspace{-3mm}
\caption{Rapidity spectra of protons for SPS energies from central collisions of Pb+Pb (NA49). Experimental data
are from the~\cite{na49.1, na49.2, na49.3, na49.4,na49.5}. Black and red histograms are DCM-SMM and UrQMD calculations, correspondingly.
}
\end{center}
\labelf{NA49_prot_y-dist}
\vspace{-5mm}
\end{figure}

\begin{figure}[ht]
\begin{center}
\includegraphics[width=100mm]{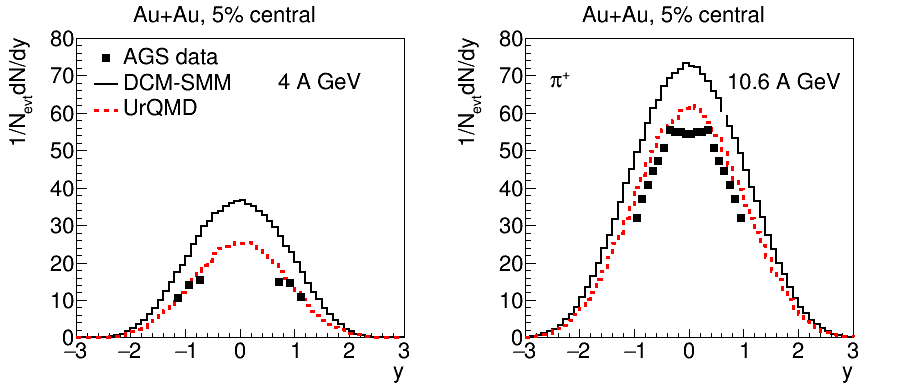}
\vspace{-3mm}
\caption{Rapidity spectra of $\pi^{+}$ in central Au+Au collisions in comparison to AGS data \cite{E0895,E877,E917,E866}. Histograms are DCM-SMM and UrQMD calculations.
}
\end{center}
\labelf{pi-rap_AGS}
\vspace{-5mm}
\end{figure}
\begin{figure}[ht]
\begin{center}
\includegraphics[width=98mm]{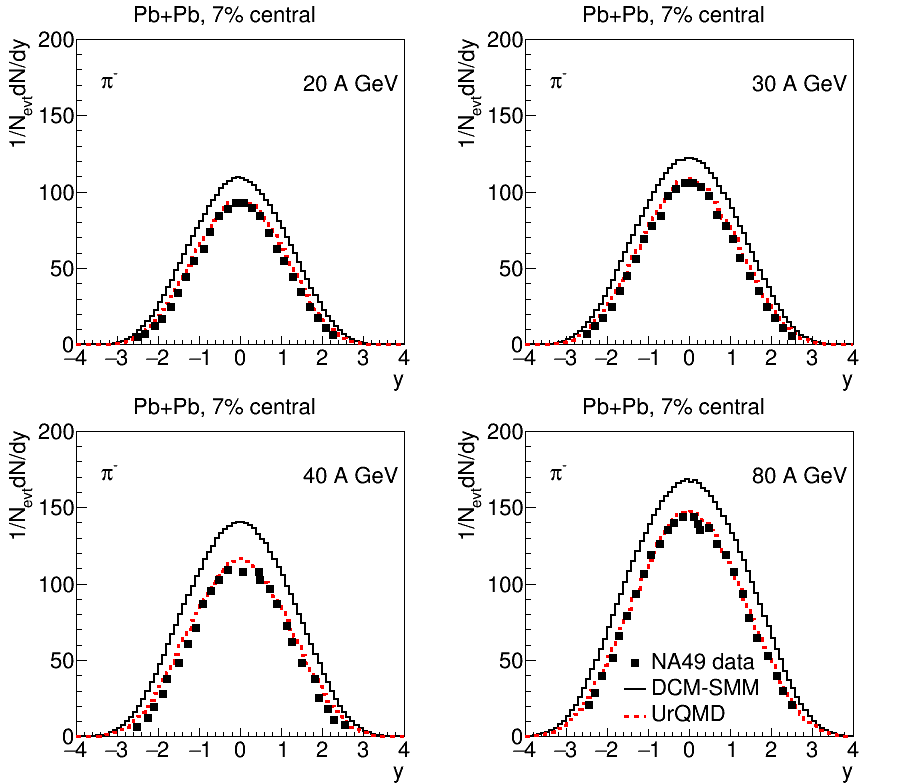}
\vspace{-3mm}
\caption{Rapidity spectra of pions for SPS energies from central Pb+Pb collisions in comparison to NA49 data  \cite{na49.1,na49.2,na49.3,na49.4,na49.5}. Black and red histograms are DCM-SMM and UrQMD calculations, correspondingly.
}
\end{center}
\labelf{pi-rap_NA49}
\vspace{-5mm}
\end{figure}
\begin{figure}[ht]
\begin{center}
\includegraphics[width=98mm]{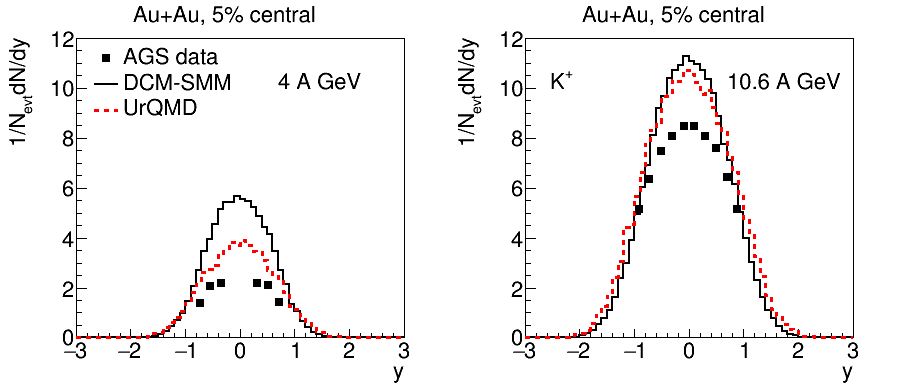}
\vspace{-3mm}
\caption{Rapidity spectra of $K^{+}$ for AGS energies from central Au+Au collisions in comparison to AGS data  \cite{E0895,E877,E917,E866}. Histograms are DCM-SMM and UrQMD calculations.
}
\end{center}
\labelf{K-rap_AGS}
\vspace{-5mm}
\end{figure}
\begin{figure}[ht]
\begin{center}
\includegraphics[width=98mm]{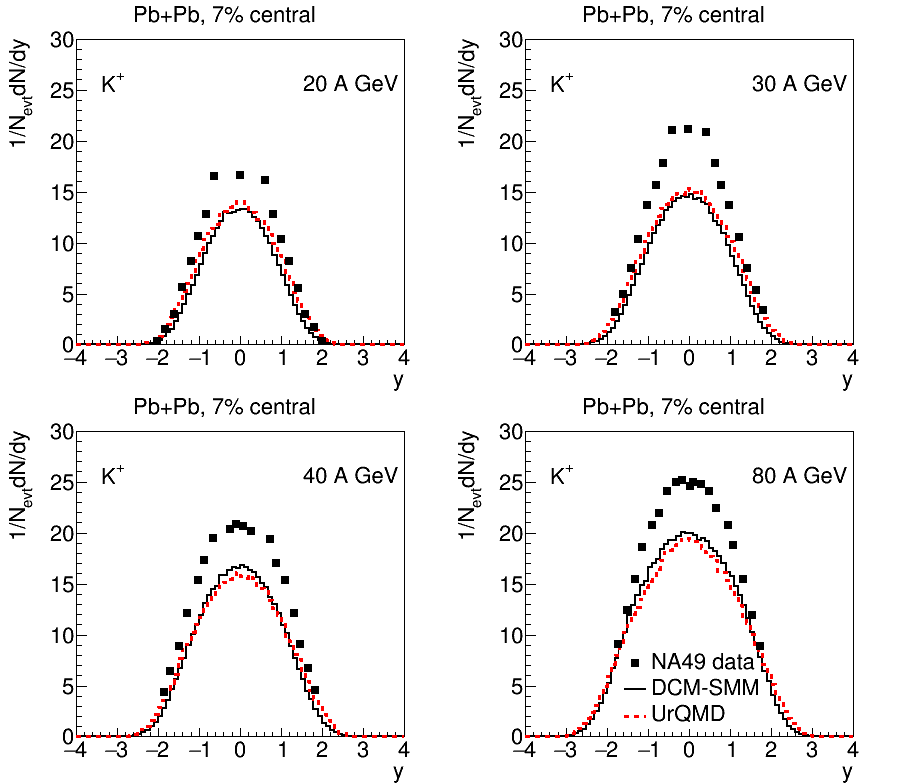}
\vspace{-3mm}
\caption{Rapidity spectra of $K^{+}$ for SPS energies from central Pb+Pb collisions in comparison to NA49 data  \cite{na49.1,na49.2,na49.3,na49.4,na49.5}. Black and red histograms are DCM-SMM and UrQMD calculations, correspondingly.
}
\end{center}
\labelf{K-rap_NA49}
\vspace{-5mm}
\end{figure}
\begin{figure}[ht]
\begin{center}
\includegraphics[width=98mm]{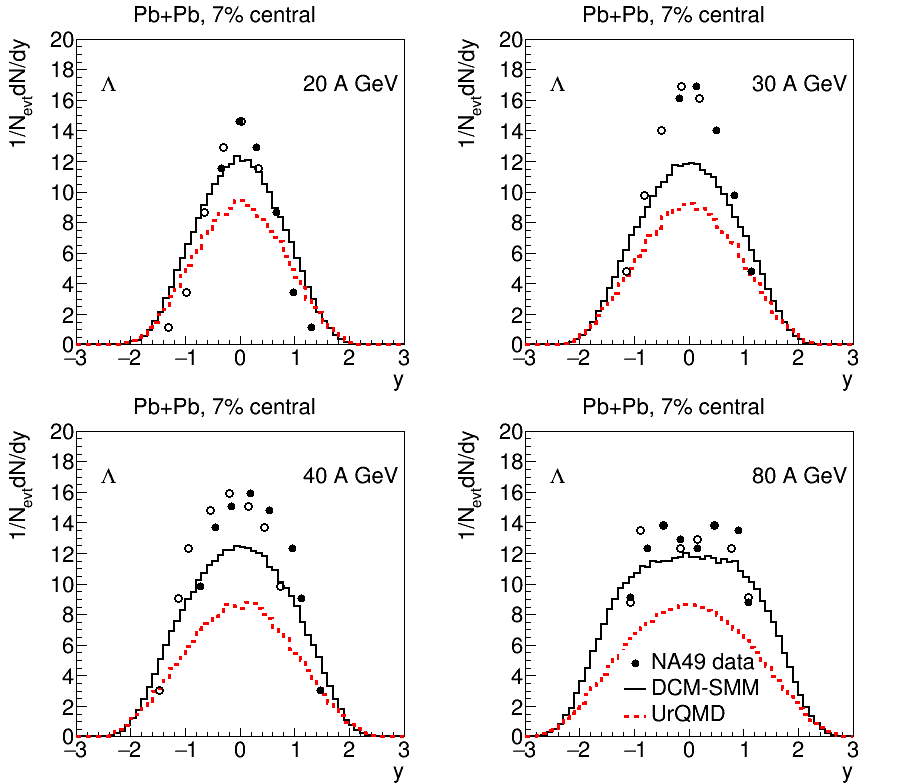}
\vspace{-3mm}
\caption{Rapidity spectra of Lambda for SPS energies from central Pb+Pb collisions in comparison to NA49 data  \cite{na49.1,na49.2,na49.3,na49.4,na49.5}. Black and red histograms are DCM-SMM and UrQMD calculations, correspondingly.
}
\end{center}
\labelf{Lam-rap_NA49}
\vspace{-5mm}
\end{figure}

\begin{figure}[ht]
\begin{center}
\begin{tabular}{cc}
\includegraphics[width=60mm]{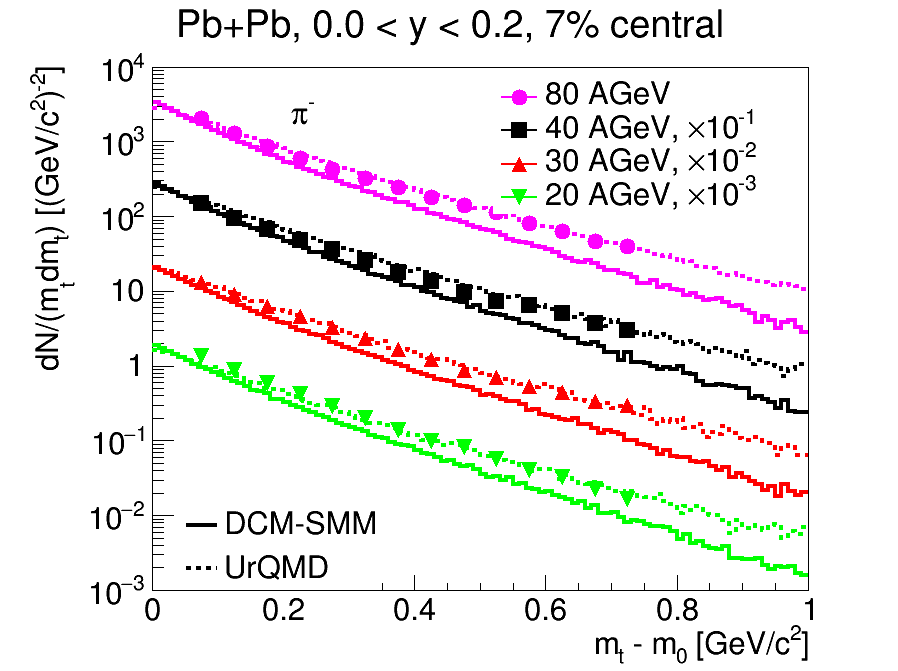}
&
\includegraphics[width=60mm]{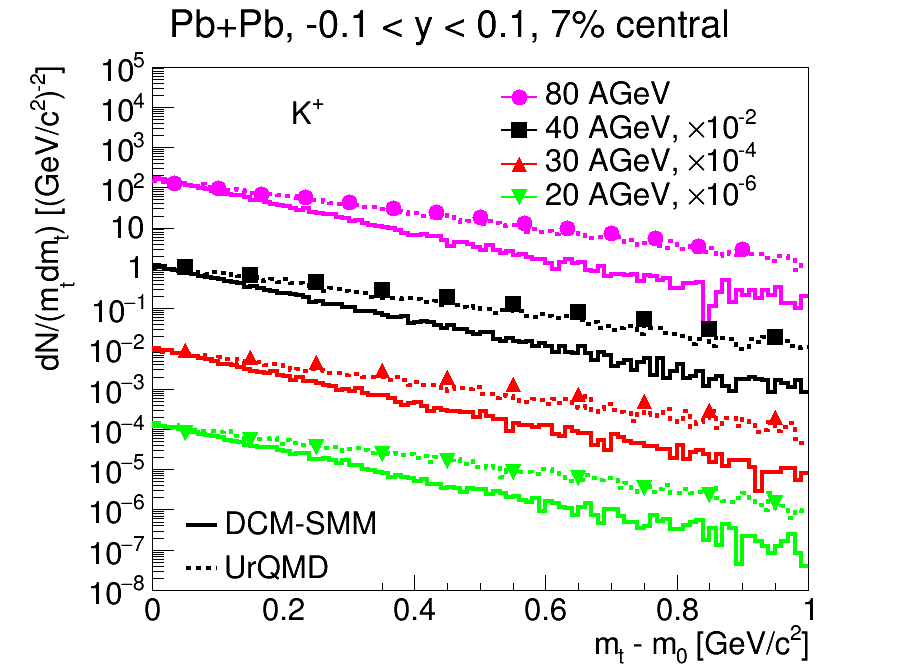}
\end{tabular}
\caption{Transverse mass distributions of $\pi^{+}$ and $K^{+}$ in central Pb+Pb collisions at NA49 energies. The data are from NA49 \cite{na49.6, na49.7, na49.8, na49.9, na49.10, na49.11,na49.12, na49.13}.}
\end{center}
\labelf{mt-dist}
\end{figure}


\begin{thebibliography}{}
%
%
\bibitem{barash} V.S.~Barashenkov, V.D.~Toneev, Interaction of high energy particles and nuclei with nuclei. // Atomizdat, 1973.
\bibitem{toneev83} \emph{Toneev V. D., Gudima K. K.}
Particle Emission In Light And Heavy Ion Reactions //
Nucl. Phys. A. 1983. V. 400, P. 173-190.

\bibitem{toneev90} \emph{Toneev V.D., Amelin N. S., Gudima K. K., Sivoklokov S. Yu.}
Dynamics of relativistic heavy ion collisions //
Nucl. Phys. A. 1990. V. 519, P. 463-478.

\bibitem{amelin51a} \emph{Amelin N. S., Gudima K. K., Toneev V. D.}
Quark - Gluon String Model and Ultrarelativistic Heavy Ion Interactions //
Sov. J. Nucl. Phys. 1990. V.51, P. 327-333.

\bibitem{amelin51b} \emph{Amelin N. S., Gudima K. K., Toneev V. D.}
Ultrarelativistic nucleus-nucleus collisions within a dynamical model of
independent quark - gluon strings //
Sov. J. Nucl. Phys. 1990. V. 51, P. 1093-1101.

\bibitem{amelin52}  \emph{Amelin N. S., Gudima K. K., Sivoklokov S. Yu., Toneev V. D.}
Further Development of a Quark - Gluon String Model for Describing High-energy
Collisions With a Nuclear Target //
Sov. J. Nucl. Phys. 1991. V.52, P. 172-178.

\bibitem{SMM-3} \emph{Bondorf J. P., Botvina A. S., Iljinov A. S., Mishustin I. N., Sneppen K.}
Statistical multifragmentation of nuclei //
Phys. Rep. 1995. V. 257, P. 133-221.

\bibitem{blann} \emph{M. Blann} // Ann. Rev. Nucl. Sci. 25 (1975) 123.

\bibitem{GEM2}  \emph{Furihata S.}
Statistical Analysis of Light Fragment Production from Medium Energy
Proton-Induced Reactions //
Nucl. Instr. Meth. B. 2000. V. 171 P. 251-258.
Development of a Generalized Evaporation Model and Study of Residual
nuclei Production //
Ph. D. Thesis, Tohoku University, Sendai, Japan, 2003.

\bibitem{qgsm} \emph{Amelin N. S., Bravina L. V}
The Monte Carlo Realization of Quark - Gluon String Model for
Description of High-Energy Hadron Hadron Interactions //
Sov. J. Nucl. Phys. 1990. V. 51, P, 133-140.
\emph{Amelin N.S., Bravina L. V., Csernai L. P., Toneev V. D., Gudima K. K., Sivoklokov S. Yu.}
Strangeness production in proton and heavy ion collisions at 200-A-GeV //
Phys. Rev. C. 1993. V. 47, 2299-2307.


\bibitem{capella} \emph{Capella A., Sukhatme U., Tran J.}
Soft multihadron production from partonic structure and fragmentation functions //
Z. Physik. 1980. V. 3, P. 329-337.


\bibitem{kaidalov} \emph{Kaidalov A. B.}
Quark and diquark fragmentation functions in the model of quark gluon strings //Sov. J. Nucl. Phys. 1987. V. 45, P. 902-907.

\bibitem{field78} \emph{Field R. D., Feynman R. P.}
A Parametrization of the Properties of Quark Jets //
Nucl. Phys. B. 1978. V. 136, P. 1-76 (1978).

\bibitem{Gudima:83a} \emph{Schulz H., R\"opke G., Gudima K. K., Toneev V. D.}
The Coalescence Phenomenon and the Pauli Quenching in High-Energy
Heavy-Ion Collisions //
Phys. Lett. B. 1983 V. 124, P. 458-460.

\bibitem{Gudima2012} \emph{Steinheimer J., Gudima K., Botvina A., Mishustin I.,
Bleicher M., St\"ocker H.}
Hypernuclei, dibaryon and antinuclei production in high energy heavy ion collisions:
Thermal production versus Coalescence //
Phys. Lett. B. 2012. V. 714, P. 85-91.

\bibitem{coal-NA49} \emph{Anticic T. et al.} (NA49 Collaboration)
Production of deuterium, tritium, and $^3$He in central
Pb+Pb collisions at 20A, 30A, 40A, 80A, and 158A GeV at the CERN SPS
Phys. Rev. B. 2016. V. 94, P. 044906.

\bibitem{Botvina2015} \emph{Botvina A. S., Steinheimer J., Bratkovskaya E., Bleicher M., Pochodzalla J.}
Formation of hypermatter and hypernuclei within transport models in relativistic ion collisions //
Phys. Lett. B. 2015. V. 742, P. 7-14.


\bibitem{Bohr} \emph{Bohr N.}
Neutron Capture and Nuclear Constitution //
Nature 1936. V. 137 P. 344-348.

\bibitem{Botvina90} \emph{Botvina A. S., Iljinov A. S., Mishustin I. N.}
Multifragment Break-up of Nuclei by Intermediate-Energy Protons //
Nucl. Phys. A. 1990. V. 507, P. 649-674.

\bibitem{Botvina94} \emph{Botvina A. S., Gudima K. K., Iljinov A. S., Mishustin I. N.}
Multifragmentation of Highly-Excited Nuclei in Nucleus-Nucleus Collisions at
Intermediate Energies //
Sov. J. Nucl. Phys. 1994. V. 57, 628-635.

\bibitem{ALADIN} \emph{Botvina A. S., Mishustin I.N., Begemann-Blaich M.,  Hubele J.,  Imme G.,  Iori I.,  Kreutz R.,  Kunde G.J., Kunze  W.D.,
Lindenstruth V., Lynen U., Moroni  A., Mialler W.F.J.,
Ogilvie C.A., Pochodzalla J., Raciti G., Rubehn Th., Sann H.,
Schtittauf A., Seidel W., Trautmann W., Werner A.},
Multifragmentation of spectators in relativistic heavy ion reactions //
Nucl. Phys. A. 1995. V. 584, P. 737-756.

\bibitem{Xi97} \emph{Xi Hongfei, Odeh T., Bassini R., Begemann-Blaich M., Botvina A. S., Fritz S., Gaff S. J., Groß C., Immé G., Iori I., Kleinevoß U., Kunde G. J., Kunze W. D., Lynen U., Maddalena V., Mahi M., Möhlenkamp T., Moroni A., Müller W. F. J., Nociforo C., Ocker B., Petruzzelli F., Pochodzalla J., Raciti G., Riccobene G., Romano F. P., Rubehn Th., Saija A., Schnittker M., Schüttauf A., Schwarz C., Seidel W., Serfling V., Sfienti C., Trautmann W., Trzcinski A., Verde G., Wörner A., Zwieglinski B.}, Breakup temperature of target spectators in Au-197 + Au-197 collisions at E/A = 1000-MeV //
Z. Phys. A. 1997. V. 359, P. 397-406.

\bibitem{Ogul11} \emph{Ogul R. et al.} (ALADIN Collaboration)
Isospin-dependent multifragmentation of relativistic projectiles //
Phys. Rev. C. 2011. V. 83, P. 024608.

\bibitem{Botvina17} \emph{Botvina A. S., Gudima K. K., Steinheimer J., Bleicher M., Pochodzalla J.}
Formation of hypernuclei in heavy-ion collisions around the threshold energies //
Phys. Rev. C. 2017. V. 95, P. 014902.

\bibitem{Eren2013} \emph{Eren N.,  Buyukcizmeci N., Ogul R., Botvina A. S.}
Mass distribution in the disintegration of heavy nuclei //
Eur. Phys. J. A, 2013. V. 49, P. 48-54.

\bibitem{jandel} \emph{Jandel M., Botvina A. S., Yennello S. J., Souliotis G. A., Shetty D. V., Bell E. and Keksis A.}
The decay time scale for highly excited nuclei as seen from asymmetrical emission of particles //
J. Phys. G. 2005 V. 31, P. 29-38.

\bibitem{XXXHFTF} \emph{Bonche P., Levit S., Vautherin D.}
Statistical properties and stability of hot nuclei //
Nucl. Phys. A. 1985. V. 436, P. 265-293; \emph{Suraud E.}
Semi-classical calculations of hot nuclei //
Nucl. Phys. A. 1987. V. 462, P. 109-149; \emph{Das Gupta S.}
Mass distributions from microscopic models of heavy ion collisions //
Phys. Rev. C. 1987. V. 35, P. 556-567; \emph{Strack B.}
Fragmentation of hot quantum drops //
Phys. Rev. C. 1987. V. 35, P. 691-695; \emph{Boal D. H., Gloshi J. N.}
From binary breakup to multifragmentation: Computer simulation //
Phys. Rev. C. 1988. V. 37, P. 91-100.

\bibitem{hubele} \emph{Hubele J. et al.}
Statistical fragmentation of Au projectiles at E/A = 600-MeV //
Phys. Rev. C. 1992. V. 46, P. 1577-1581.

\bibitem{Deses} \emph{Desesquelles P. et al.}
Global protocol for comparison of simulated data with experimental data //
Nucl. Phys. A. 1996. V. 604, P. 183-207.

\bibitem{napolit} \emph{Napolitani P. et al.}
High-resolution velocity measurements on fully identified light nuclides produced
in Fe-56 + hydrogen and Fe-56 + titanium systems //
Phys. Rev. C. 2004. V. 70, P. 054607.

\bibitem{ISIS} \emph{Pienkowski L. et al.}
Breakup time scale studied in the 8-GeV/c pi-+ Au-197 reaction //
Phys. Rev. C. 2002. V. 65, P. 064606.

\bibitem{beaulieu} \emph{Beaulieu L. et al.}
Signals for a transition from surface to bulk emission in thermal multifragmentation //
Phys. Rev. Lett. 2000. V. 84, P. 5971-5974.

\bibitem{karnaukhov} \emph{Karnaukhov V. A. et al.}
Thermal multifragmentation of hot nuclei and liquid-fog phase transition //
Phys. Atom. Nucl. 2003. V. 66, P. 1242-1251.

\bibitem{Jackson} \emph{Jackson, A.D.; Mishustin, I.; Botvina, A.S.}
Partitioning Composite Finite Systems //
Phys. Rev. E. 2000. V. 62, P. 64-67 (2000).

\bibitem{fermi} \emph{Fermi E.}
High-energy nuclear events //
Prog. Theor. Phys. 1950. V. 5, P. 570-583.

\bibitem{Botvina01} \emph{Botvina A. S., Mishustin I. N.}
Statistical evolution of isotope composition of nuclear fragments //
Phys. Rev. C. 2001. V. 63, P. 061601.

\bibitem{Botvina87} \emph{Botvina A. S. et al.}
Statistical simulation of the breakup of highly excited nuclei //
Nucl. Phys. A. 1987. V. 475, P. 663-686.

\bibitem{MSU} \emph{D'Agostino M. et al.}
Statistical multifragmentation in central Au + Au collisions at 35-MeV/u //
Phys. Lett. B. 1996. V. 371, P. 175-180.

\bibitem{INDRA} \emph{Bellaize N. et al.}
Multifragmentation process for different mass asymmetry in the entrance channel
around the Fermi energy //
Nucl. Phys. A. 2002. V. 709, P. 367-391.

\bibitem{Iglio} \emph{Iglio J. et al.}
Symmetry energy and the isoscaling properties of the fragments produced in
40-Ar, 40-Ca + 58-Fe, 58-Ni reactions at 25, 33, 45, and 53 MeV/nucleon //
Phys. Rev. C. 2006. V. 74, P. 024605.

\bibitem{Souliotis} \emph{Souliotis G. et al.}
Tracing the evolution of the symmetry energy of hot nuclear fragments from
the compound nucleus towards multifragmentation //
Phys. Rev. C. 2007. V. 75, P. 011601.

\bibitem{EOS1} \emph{Hauger J. A. et al.}
Two-stage multifragmentation of 1A GeV Kr, La, and Au //
Phys. Rev. C. 2000. V. 62, P. 024616.


\bibitem{EOS2} \emph{Scharenberg R. P. et al.}
Comparison of 1-A-GeV Au-197 + C data with thermodynamics: The Nature of phase
transition in nuclear multifragmentation  //
Phys. Rev. C. 2001. V. 64, P. 054602.

\bibitem{FAZA} \emph{Avdeyev S. P. et al.}
Comparative study of multifragmentation of gold nuclei induced by relativistic protons, He-4, and C-12 //
Nucl. Phys. A. 2002. V. 709, P. 392-414.

\bibitem{Ahmad_CAg} \emph{Ahmad T., Irfan M.}
Inelastic Interactions Caused by 4.5A GeV/c Carbon and Silicon Nuclei //
Nuov. Cim. A. 1993. V. 106, P. 171-185.

\bibitem{EOSd} \emph{Porile N. T. et al.
} (EOS Collab.),
Multifragmentation of 1-A-GeV Kr, La, and Au on carbon //
Nucl. Phys. A. 2001 V. 681, P. 253-266.

\bibitem{Botvina11} \emph{Botvina A. S., Gudima K. K., Steinheimer J., Bleicher M., Mishustin I. N}
Production of spectator hypermatter in relativistic heavy-ion collisions //
Phys. Rev. C. 2011 V. 84, P. 064904.

\bibitem{Botvina07} \emph{Botvina A. S., Pochodzalla J.}
Production of hypernuclei in multifragmentation of nuclear spectator matter //
Phys. Rev. C. 2007. V. 76, P. 024909.

\bibitem{Buyuk13} \emph{Buyukcizmeci N., Botvina A. S., Pochodzalla J., Bleicher M.}
Mechanisms for the production of hypernuclei beyond the neutron and proton drip lines //
Phys. Rev. C. 2013. V. 88, P. 014611.

\bibitem{Botvina16} \emph{Botvina A. S., Buyukcizmeci N., Ergun A., Ogul R., Bleicher M., Pochodzalla J.}
Formation of hypernuclei in evaporation and fission processes //
Phys. Rev. C. 2016. V. 94, P. 054615.

\bibitem{Botvina13} \emph{Botvina A. S., Gudima K. K., Pochodzalla J.}
Production of hypernuclei in peripheral relativistic ion collisions //
Phys. Rev. C. 2013. V. 88, P. 054605.

\bibitem{Buyuk18} Buyukcizmeci N., Botvina A. S., Ergun A., Ogul R., Bleicher M.
Statistical production and binding energy of hypernuclei //
Phys. Rev. C. 2018 V. 98, P. 064603.

\bibitem{gudtitov} \emph{Gudima K. K., Titov A. I., Toneev V. D.}
Hadronic sources of dileptons from nuclear collisions at intermediate and relativistic energies //
Phys. Lett. B. 1992. V. 287, P. 302-306.

\bibitem{nica} \emph{Blaschke D. et al.}
Topical issue on Exploring Strongly Interacting Matter at High Densities - NICA White Paper //
Eur. Phys. J. A. 2016. V. 52, P. 267.
NICA White Paper,
http://theor.jinr.ru/twiki-cgi/view/NICA/WebHome
http://nica.jinr.ru/files/WhitePaper.pdf.

\bibitem{STAR:2017ckg} \emph{Adamczyk L. et al.} (STAR Collaboration)
Global $\Lambda$ hyperon polarization in nuclear collisions //
Nature. 2017. V. 548, P. 62.

\bibitem{PhysRevC.98.014910} \emph{J. Adam et al.} (STAR Collaboration)
Global polarization of $\Lambda$ hyperons in Au+Au collisions at $\sqrt{s_{NN}} = 200 GeV$ //
Phys. Rev. C. 2018. V. 98, P.014910.

\bibitem{Betz:2007kg} \emph{Betz B., Gyulassy M., Torrieri G.}
Polarization probes of vorticity in heavy ion collisions //
Phys. Rev. C. 2007. V. 76, P. 044901.

\bibitem{Baznat:2013zx} \emph{Baznat M., Gudima K., Sorin A., Teryaev O.}
Helicity separation in heavy-ion collisions //
Phys. Rev. C88, 061901 (2013), arXiv:1301.7003 [nucl-th].

\bibitem{Baznat:2015eca} \emph{Baznat M. I., Gudima K. K., Sorin A. S., Teryaev O. V.}
Femto-vortex sheets and hyperon polarization in heavy-ion collisions //
Phys. Rev. C. 2016. V.93, P. 031902.

\bibitem{Teryaev:2015gxa} \emph{Teryaev O., Usubov R.}
Vorticity and hydrodynamic helicity in heavy-ion collisions in the in the
hadron-string dynamics model //
Phys. Rev. C. 2015. V. 92, P. 014906.

\bibitem{Sorin:2016smp} \emph{SorinA., Teryaev O.}
Axial anomaly and energy dependence of hyperon polarization in heavy-ion collisions //
Phys. Rev. C. 2017. V. 95, No. 1. P. 011902.

\bibitem{Becattini:2013fla} \emph{Becattini F., Chandra V., Del Zanna L., Grossi E.}
Relativistic distribution function for particles with spin at local thermodynamical equilibrium //
Annals. Phys. 2013. V. 338, P. 32-49.

\bibitem{Fang:2016vpj} \emph{Fang R.-H., Pang L.-G., Wang Q., Wang X.-N.}
arXiv 1604.04036. 2016.

\bibitem{STAR_Lam-pol} \emph{L. Adamczyk et al.} (STAR Collaboration) //
Nature 2017. V. 548, P. 62.

\bibitem{urqmd-3.4} The UrQMD Model // http:\\urqmd.org

\bibitem{fritiof} \emph{Andersson B., Gustafson G, Nilsson-Almqvist B.}
A Model for Low p(t) Hadronic Reactions, with Generalizations to Hadron - Nucleus
and Nucleus-Nucleus Collisions //
Nucl. Phys. B. 1987. V. 281, P. 289-309.

\bibitem{grt} \emph{Gribov V.}
A Reggeon Diagram Technique //
Sov. JETP. 1968. V. 26, P. 414-423 (1968)


\emph{Gribov L. V., Levin E. M., Ryskin M. G.}
Semihard Processes in QCD //
Phys. Rep. 1983. V. 100, P. 1-150.

\bibitem{PaSh07} \emph{Pajares C., Shabelski Yu. M.}
Relativistic Nuclear Interactions //
M.: URSS, 2007. 272 c.


\emph{Kaidalov A. B.}
Soft interactions of hadrons in QCD //
Surveys in High Energ. Phys. 1999. V. 13, P. 265-330.

\bibitem{jetset} \emph{Sj\"ostrand T.}
The Lund Monte Carlo for Jet Fragmentation and e+ e- Physics: Jetset Version 6.2 //
Comp. Phys. Comm. 1986. V. 39, P. 347-407.

\bibitem{bravinac78} \emph{Bravina L. A.
, I.~Arsene, M.S.~Nilsson, E.E.~Zabrodin, J.~Bleibel, Amand Faessler, C.~Fuchs, M.~Bleicher, G.~Burau, H.Stocker},
Microscopic models and effective equation of state in nuclear collisions at FAIR energies //
Phys. Rev. C. 2007. V. 78, P. 014907.

\bibitem{E0895} \emph{Klay J. L. et al.} (E-0895 Collaboration),
Charged Pion Production in 2 to 8 AGeV Central Au+ Au Collisions //
Phys. Rev. C. 2003. V. 68, P. 054905.

\bibitem{E877} \emph{Barrette J. et al.} (E877 Collaboration),
Proton and Pion Production in Au+ Au Collisions at 10.8 A GeV/c //
Phys. Rev. C. 2000. V. 62, P. 024901.

\bibitem{E917} \emph{Back B. B. et al.} (E917 Collaboration),
Baryon Rapidity Loss in Relativistic A u+ A u Collisions //
Phys. Rev. Lett. 2001. V. 86, P. 1970-1973.

\bibitem{E866} \emph{Stachel J.}
Towards the quark-gluon-plasma //
Nucl. Phys. A. 1999. V. 654, P. 119-135.

\bibitem{na49.1} \emph{Appelshauser H. et al.} (NA49 Collaboration),
Baryon Stopping and Charged Particle Distributions in Central Pb+Pb
Collisions at 158 GeV per Nucleon  //
Phys. Rev. Lett. 1999. V. 82, P. 2471-2475.

\bibitem{na49.2} \emph{Anticic T. et al.} 
Energy and centrality dependence of deuteron and proton production
in Pb + Pb collisions at relativistic energies //
Phys. Rev. C. 2004. V. 69, P. 024902.

\bibitem{na49.3} \emph{Alt C. et al.} 
Energy dependence of particle ratio fluctuations in central Pb+ Pb
collisions from $\sqrt{s_{NN}}$=6.3 to 17.3 GeV //
Phys. Rev. C. 2006. V. 73, P. 044910.

\bibitem{na49.4} \emph{Blume C. et al.} 
Centrality and energy dependence of proton, light fragment and hyperon
production //
J. Phys. G. 2007. V. 34, P. S951-954.

\bibitem{na49.5} \emph{Anticic T. et al.} 
Centrality dependence of proton and antiproton spectra in Pb+Pb
collisions at 40A GeV and 158A GeV measured at the CERN Super
Proton Synchrotron //
Phys. Rev. C. 2011. V. 83, P. 014901.

\bibitem{E895.2} \emph{Pinkenburg C. et al.} 
Production and collective behavior of strange particles in Au + Au
collisions at 2-AGeV - 8-AGeV //
Nucl. Phys. A. 2002. V. 698, P. 495-498. 

\bibitem{E895.3} \emph{Chung P. et al.} 
Near-Threshold Production of the Multistrange $\Xi$-Hyperon //
Phys. Rev. Lett. 2003. V. 91, P. 202301. 

\bibitem{na49.6} \emph{Alt C. et al.} 
Pion and kaon production in central Pb + Pb collisions at 20-A and
30-A-GeV: Evidence for the onset of deconfinement //
Phys. Rev. C. 2008. V. 77, P. 024903. 

\bibitem{na49.7} \emph{Afanasiev S. V. et al.} 
Energy dependence of pion and kaon production in central Pb + Pb
collisions //
Phys. Rev. C. 2002. V. 66, P. 054902. 

\bibitem{na49.8} \emph{Anticic T. et al.} 
$\Lambda$ and $\overline{\Lambda}$ Production in Central Pb-Pb Collisions
at 40, 80, and 158A GeV //
Phys. Rev. Lett. 2004. V. 93, P. 022302. 

\bibitem{na49.9} \emph{Richard A. ey al.} 
Energy dependence of hyperon production in central Pb + Pb collisions
at the CERN-SPS //
J. Phys. G. 2005. V. 31, P. S155-S162.

\bibitem{na49.10} \emph{Mitrovski M. K. et al.} 
Strangeness production at SPS energies //
J. Phys. G. 2006. V. 32, P. S43-S50. 

\bibitem{na49.11} \emph{Blume C. et al.} 
Review of results from the NA49 collaboration //
J. Phys. G. 2005. V. 31, P. S685-S692. 

\bibitem{na49.12} \emph{Afanasiev S. V. et al.} 
Cascade and anti-Cascade+ production in central Pb + Pb collisions
at 158-GeV/c per nucleon //
Phys. Lett. B. 2002. V. 538, P. 275-281.  

\bibitem{na49.13} \emph{Alt C. et al.} 
Omega- and anti-Omega+ production in central Pb + Pb collisions
at 40-AGeV and 158-AGeV //
Phys. Rev. Lett. 2005. V. 94, P. 192301. 

\bibitem{urqmd-2.3} \emph{Petersen H., Bleicher M., Bass S.A., Stoker H.}
UrQMD-2.3 - Changes and Comparisons // [arXiv:hep-ph/0805,0567].

\end{thebibliography}
\end{document}